\documentclass[epj]{svjour}

\usepackage{graphics}
\usepackage{amsmath}
\usepackage{amssymb}
\usepackage{multirow}
\usepackage{upgreek}
\usepackage[switch]{lineno}
\usepackage{color}
\usepackage{colortbl}
\usepackage{blindtext}
\usepackage{enumerate}
\usepackage{setspace}

\usepackage[utf8]{inputenc}
\newcommand\correspondingauthor{\thanks{Corresponding author: pawel\_piotr.staszel@uj.edu.pl.}}

\newcommand{\eV}{\ensuremath{\mbox{e\kern-0.1em V}}\xspace}
\newcommand{\GeV}{\ensuremath{\mbox{Ge\kern-0.1em V}}\xspace}
\newcommand{\MeV}{\ensuremath{\mbox{Me\kern-0.1em V}}\xspace}
\newcommand{\GeVc}{\ensuremath{\mbox{Ge\kern-0.1em V}\!/\!c}\xspace}
\newcommand{\GeVcc}{\ensuremath{\mbox{Ge\kern-0.1em V}\!/\!c^2}\xspace}
\newcommand{\MeVcc}{\ensuremath{\mbox{Me\kern-0.1em V}\!/\!c^2}\xspace}
\newcommand{\AGeV}{\ensuremath{A\,\mbox{Ge\kern-0.1em V}}\xspace}
\newcommand{\AGeVc}{\ensuremath{A\,\mbox{Ge\kern-0.1em V}\!/\!c}\xspace}
\newcommand{\MeVc}{\ensuremath{\mbox{Me\kern-0.1em V}/c}\xspace}

\newcommand{\dd}{\ensuremath{{\textrm d}}\xspace}
\newcommand{\dedx}{\ensuremath{\dd E\!~/~\!\dd x}\xspace}

\newcommand{\pt}{\ensuremath{p_{\textrm T}}\xspace}

\newcommand{\GeantFour}{{\scshape Geant4}\xspace}

\newcommand{\CernVM}{\textsc{Cern\-\kern-0.05emVM}\xspace}

\newcommand{\TeV}{\ensuremath{\mbox{Te\kern-0.1em V}}\xspace}

\newcommand{\NASixtyOne}{NA61\slash SHINE\xspace}
\newcommand{\coordinate}[1]{{\fontfamily{lmss}\selectfont#1}}
\newcommand{\mcoor}[1]{\text{{\fontfamily{lmss}\selectfont#1}}}
\newcommand{\pl}{$p_{\text{L}}$}

\graphicspath{{plots/}}

\begin{document}
 \twocolumn


\title{A high-resolution pixel silicon Vertex Detector for open charm measurements
with the \NASixtyOne spectrometer at the CERN SPS}
\author{
A.~Aduszkiewicz\inst{1}\and
M.~Bajda\inst{2}\and
M.~Baszczyk\inst{3}\and
W.~Bryliński\inst{4}\and
J.~Brzychczyk\inst{2}\and
M.~Deveaux\inst{5, 12}\and
P.~Dorosz\inst{3} \and
S.~Di Luise\inst{6}\and
G.~Feofilov\inst{7}\and
M.~Gazdzicki\inst{5, 8}\and
S.~Igolkin\inst{7}\and
M.~Jabłoński\inst{2}\and
V.~Kovalenko\inst{7}\and
M.~Koziel\inst{5}\and
W.~Kucewicz\inst{3}\and
D.~Larsen\inst{2}\and
T.~Lazareva\inst{7}\and
K.~Łojek\inst{2}\and
Z.~Majka\inst{2}\and
P.~Martinengo\inst{9}\and
A.~Merzlaya\inst{2}\and
L.~Mik\inst{3,11}\and
R.~Płaneta\inst{2}\and
P.~Staszel\inst{2}\correspondingauthor\and
M.~Suljic\inst{10}\and
D.~Tefelski\inst{4}\and
V.~Vechernin\inst{7}
}

\institute{
University of Warsaw, Warsaw, Poland\and 
Jagiellonian University in Krakow, Krakow, Poland\and
AGH University of Science and Technology, Krakow, Poland\and
Warsaw University of Technology, Warsaw, Poland\and
Goethe University Frankfurt, Frankfurt, Germany\and
Eidgenössische Technische Hochschule Zürich, Zürich, Switzerland\and
St. Petersburg State University, St. Petersburg, Russia\and
Jan Kochanowski University, Kielce Poland\and
European Organization for Nuclear Research (CERN), Geneva, Switzerland\and
University \& INFN Trieste, Trieste, Italy\and
University of Applied Sciences in Tarnow, Tarnow, Poland\and
GSI Helmholtzzentrum für Schwerionenforschung, Darmstadt, Germany
}

\date{Received: date / Revised version: date}

\abstract{
The study of open charm meson production provides an efficient tool for the investigation of the properties 
of hot and dense matter formed in nucleus-nucleus collisions. The interpretation of the existing di-muon 
data from the CERN SPS suffers from a lack of knowledge on the mechanism and properties of the open charm 
particle production. Due to this, the heavy-ion programme of the \NASixtyOne experiment at the CERN SPS 
has been extended by precise measurements of charm hadrons with short lifetimes.
A new Vertex Detector for measurements of the rare processes of open charm production in nucleus-nucleus 
collisions was designed to meet the challenges of track registration and high resolution 
in primary and secondary vertex reconstruction.
A small-acceptance version of the vertex detector was installed in 2016 and tested with Pb+Pb collisions 
at 150\AGeVc. It was also operating during the physics data taking on Xe+La and Pb+Pb collisions 
at 150\AGeVc conducted in 2017 and 2018. 
This paper presents the detector design and construction, data calibration, event reconstruction, 
and analysis procedure. 
}

\authorrunning{\NASixtyOne Collaboration}
\titlerunning{High-resolution silicon Vertex Detector for open charm measurements at the CERN SPS}
\maketitle

\section{Introduction}
\label{sec:1}


The charm production mechanism is one of the important questions in relativistic heavy-ion physics.
Several models were introduced to describe charm production.
Some are based on dynamical and others - on statistical approaches. 
Predictions of these models on the mean number of produced $c\bar{c}$ pairs ($\langle c\bar{c}\rangle$) for
central Pb+Pb collisions at 158\AGeVc differ by up to a factor of 50 \cite{Brock:1993sz,Bravina:1999dh}.
Moreover, the system size dependence is different in these approaches and the
the predictions suffer from large systematic uncertainties \cite{Gazdzicki:1998vd,NA61Proposal}.
Precise data on $\langle c\bar{c}\rangle$ will allow to 
disentangle between theoretical predictions and learn about the charm quark and hadron production mechanism.
Obtaining good estimate of $\langle c\bar{c}\rangle$ requires measurements of $D^0$, $D^+$ and their antiparticles.
This is because these 
mesons carry about 85\% of the total produced charm in Pb+Pb collisions at the top SPS energy~\cite{Cassing:2009vt,Bratkowskaya}.

Besides this, a study
of open charm meson production was proposed as a sensitive tool for detailed investigations of 
the properties of hot and dense matter formed in nucleus-nucleus collisions at ultra-relativistic 
energies \cite{Matsui:1986dk,Satz:2013ama,Muller:2013dea}. 
In particular, charm mesons are of vivid interest when studying the phase transition between confined hadronic 
matter and the quark-gluon plasma (QGP).
The $ c\bar{c}$ pairs produced in the collisions are converted into open charm mesons and charmonia ($J/\psi$ mesons and their excited states).
The charm production is expected to be different in the confined and deconfined matter because 
of the different properties of charm carriers in these phases. In confined matter, the lightest 
charm carriers are $D$ \linebreak
mesons, whereas, in deconfined matter, the carriers are charm quarks. 
The production of a $D\overline{D}$ pair ($2 \text{m}_D = 3.7$ GeV) requires more energy 
than the production of a $c\bar{c}$
pair ($2 \text{m}_c = 2.6$ GeV). Since the effective degrees of freedom of charm, hadrons and charm quarks 
are similar~\cite{Poberezhnyuk:2017ywa}, more abundant charm production is expected in deconfined than confined matter. 
Consequently, in analogy to strangeness~\cite{Gazdzicki:1998vd,Rafelski:1982pu}, a change in collision energy 
dependence of $\langle c\bar{c}\rangle$ production may 
indicate an onset of deconfinement.  

Finally, systematic measurements of open charm production are urgently needed to interpret existing results on $J/\psi$.
Such measurements would allow disentangling between initial and final state effects, revealing hidden and open 
charm transport properties  through the dense medium created in nucleus-nucleus collisions and testing the 
validity of theoretical models~\cite{Satz:2013ama}.
	
The \NASixtyOne experiment plans to measure open charm production in heavy-ion collisions in full-phase space at the SPS energies.
To observe the energy dependence of open charm production, 
additional corresponding 
measurements at higher (LHC \cite{Meninno:2017ezl,Hou:2016hgs,Simko:2017jtj,Nagashima:2017mpm}, 
RHIC \cite{Odyniec:2013kna,Yang:2017llt,Meehan:2016qon}) and lower 
(FAIR \cite{Friman:2011zz}, J-PARC \cite{Sako:2016edz}, NICA \cite{Kekelidze:2017tgp}) energies are 
needed.
	
Measurements of open charm mesons are challenging since the yields of $D$ mesons are low, and their lifetimes are 
relatively short ($c\tau=122~\upmu$m). The measurements require precise tracking and high primary and secondary vertex resolutions.
To meet these challenges, a novel high-resolution Small Acceptance Vertex Detector (SAVD) was designed and built 
under the leadership of the Jagiellonian University group participating in the \NASixtyOne experiment.
SAVD was installed as a part of the \NASixtyOne facility in December 2016.
Test data on Pb+Pb collisions at 150\AGeVc beam momenta were collected and analyzed. 
The main goal of the test was to prove the feasibility of precise tracking in the large track multiplicity 
environment and demonstrate the ability of precise primary and secondary vertex reconstruction.
In 2017 and 2018, data on  Xe+La and Pb+Pb collisions at the beam momenta of 150\AGeVc were 
recorded with SAVD included in the detector setup.
The data quality and statistics were sufficient 
for the first direct observation of a $D^0 + \overline{D^0}$ signal in the $\pi + K$ decay 
channel in nucleus-nucleus collisions at the SPS energy.
This paper presents SAVD design and construction, data calibration, event reconstruction, and analysis procedure. 



It is foreseen that the \NASixtyOne Collaboration will perform large statistics measurements after 2022. These data 
will allow for the first insight into the centrality dependence of open charm~\cite{NA61Proposal}.

The following variables and definitions are used in this paper. The particle rapidity $y$ is calculated 
in the nucleon-nucleon collision center of the mass system (c.m.s.) with 
\[
y=0.5\ln{[(E+p_{\text{L}})/(E-p_{\text{L}})]}~,
\]
where $E$ and \pl~are the particle energy and longitudinal momentum, respectively, the transverse  
momentum is denoted as \pt,~and $m$ is the particle mass. The quantities are given either in \GeV or  in \MeV.
The results shown in this paper were obtained for Xe+La collisions at the beam momenta of 150\AGeVc.

\section{\NASixtyOne experimental facility}
\label{sec:2}

\begin{figure*}
\resizebox{\linewidth}{!}{
\includegraphics{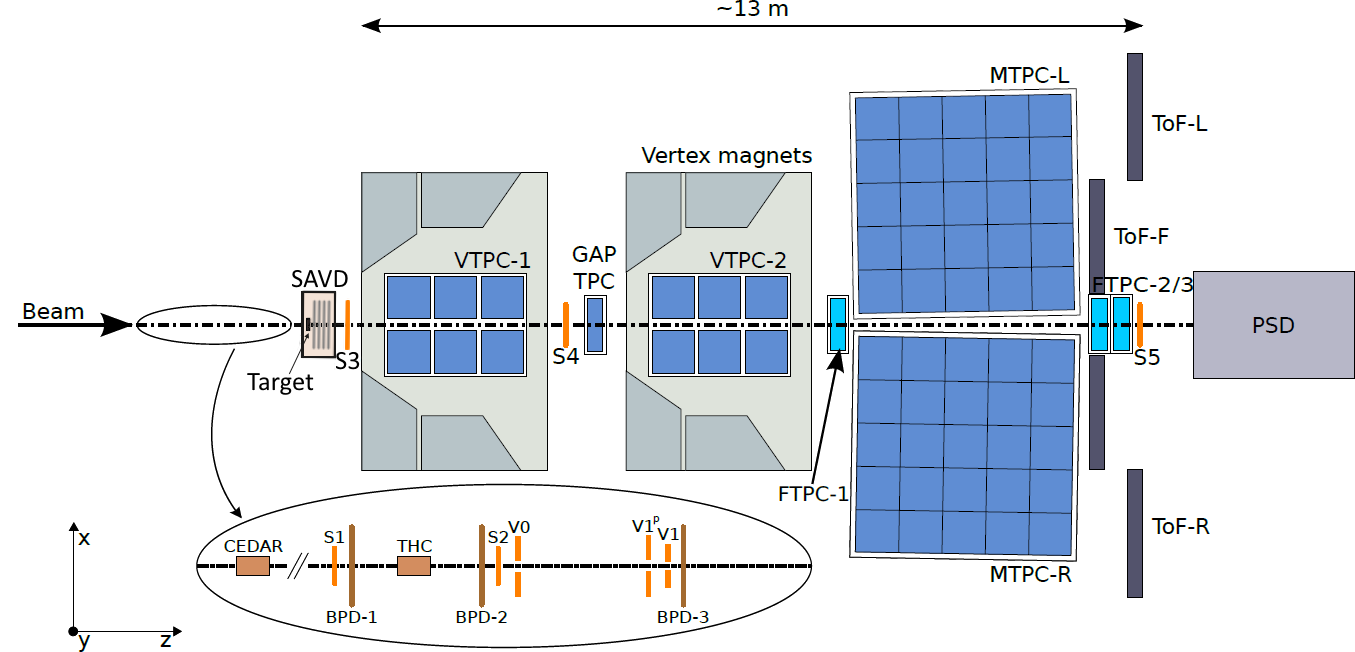}
}
\caption{The layout of the \NASixtyOne experimental setup (top view, not to scale).}
\label{fig:Figure5}
\end{figure*}

The SPS Heavy Ion and Neutrino Experiment (\NASixtyOne)~\cite{Abgrall:2014xwa} at CERN was designed to study the 
properties of the onset of deconfinement and search for the critical point of the strongly interacting matter. 
These goals are being pursued by investigating p+p, p+A and A+A collisions at different beam momenta from 13$A$ to 158\AGeVc 
for ions and up to 400~\GeVc for protons.

The layout of the experimental setup is shown in Fig.~\ref{fig:Figure5}. 
The setup includes the beam position detectors (BPD), Cherenkov counters and the scintillator detectors located upstream of the target.
They provide information on the timing, charge and position of beam particles. 
Further, the experiment includes two Vertex Time Projection Chambers (VTPC-1 and VTPC-2) located inside the vertex magnets, 
two main TPCs (MTPC-L and MTPC-R) for \dedx measurements and Gap TPC and Forward TPCs that complete the coverage between MTPCs.
These TPCs provide acceptance in the full forward hemisphere, down to $p_T$ = 0.
The TPCs allow tracking, momentum and charge determination, and measuring the mean energy loss per unit path length. 
The time-of-flight (ToF) walls used for additional particle identification are located behind the main TPCs. 
The projectile spectator detector (PSD) measures the energy of the projectile spectator and  delivers information on the collision centrality.

\subsection{Vertex Detector rationale}
\label{sec:2.1}

\begin{figure*}
\resizebox{\linewidth}{!}{
\includegraphics{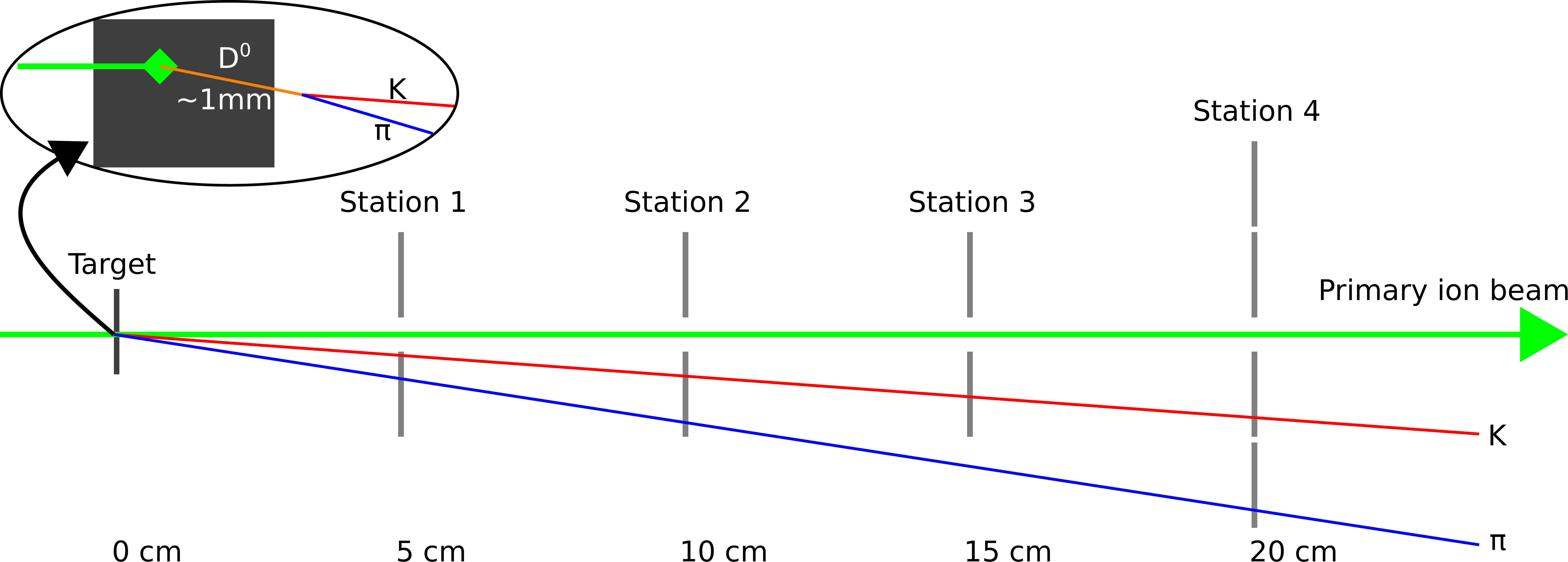}
}
\caption{Schematics of reconstruction strategy of $D^0 \rightarrow \pi^+ + K^-$ decay channel with the help of the Vertex Detector.}
\label{fig:Figure6}
\end{figure*}

For open charm measurements in nucleus-nucleus collisions, \NASixtyOne was upgraded with SAVD. 
As was already mentioned, open charm mesons are difficult to measure because of their low yields and short lifetime.
They can be 
measured in their decay channels into pions and kaons. However, in heavy-ion collisions, pions and 
kaons are produced in large numbers in other processes giving huge combinatorial background.
To distinguish the daughter particles of $D^0$ mesons from hadrons produced directly in the nucleus-nucleus interaction, 
one  selects  hadron pairs created in secondary  vertices.
The vertex reconstruction is done by extrapolating the track trajectories back to the target and identifying intersection points. 
The primary vertex will appear as the intersection point of multiple tracks while the tracks originating from selected 
decays will intersect at the 
displaced point (secondary vertex), see Fig.~\ref{fig:Figure6}.

Until the development of silicon sensors for particle tracking, it was not possible to perform secondary vertex 
reconstruction with resolution sufficient to measure open charm.
Consequently, an open charm meson production study has never been performed at the SPS energy range.
The construction of SAVD opened up  the possibility of open charm measurements.

\section{SAVD hardware}
\label{sec:3}

\begin{figure}
\resizebox{\linewidth}{!}{
\includegraphics{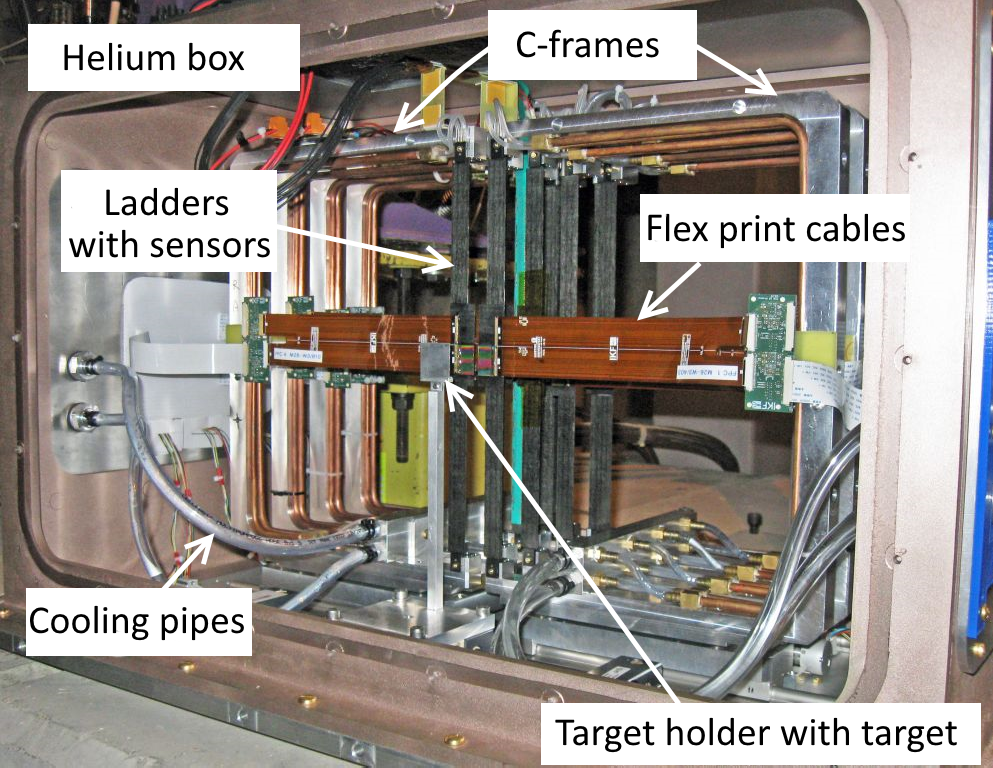}
}
\caption{Photograph of SAVD before closing the detector with the front and exit windows. The detector elements are indicated. For detail, see the text.}
\label{fig:Figure8}
\end{figure}

SAVD is positioned between the target and VTPC-1 (see Fig.~\ref{fig:Figure5}) in the in-homogeneous and weak (0.13 - 0.25T) 
fringe field of the VTPC-1 magnet. 
A photograph of the device is shown in Fig.~\ref{fig:Figure8}. 
It consists of two arms called Jura and Saleve arm. This naming follows the \NASixtyOne convention for the left and right partition of the experiment in the direction of the beam, respectively,
and  
corresponds to the location of the nearby mountains. SAVD is composed of four detection 
planes (stations) equipped with the position-sensitive MIMOSA-26AHR CMOS Monolithic Active Pixel
Sensors (MAPS)~\cite{Baudot:2009,Hu:2010,Deveaux:2011zz} provided by the PICSEL group of the IPHC Strasbourg. 
The arms are horizontally movable, allowing the sensors to be placed safely during beam tuning.
The stations, called Vds1, Vds2, Vds3 and Vds4, are located 5, 10, 15 and 20$~\rm cm$ downstream the target, respectively. 
The sensors are held and water-cooled by vertically oriented ALICE ITS carbon fibre support ``ladders''~\cite{Abelevetal:2014dna} 
developed by St. Petersburg State University and CERN.
The ladders are mounted in C-frames made from aluminum. The four C-frames of each arm share a movable support plate.
The first (Vds1) and second station (Vds2) consist of two ladders, each holding one sensor only, the third station consists of two ladders, each holding two sensors, and the last station is composed of four ladders, each hosting two sensors (see Fig.~\ref{fig:sensorsNames}).
A holder for targets is placed on an additional, movable support.  

The whole structure is installed on a thick aluminum base plate, which provides mechanical stability. 
Four brass screws serve as legs for the plate and enable fine adjustment of the vertical position when installed on the beam-line.
The pink color box structure in the photograph is made of plexiglass covered with conducting paint.
The base plate, together with the plexiglass structure and front and back mylar windows (dismounted on the photograph) 
served as a gas-tight detector box.   
During data taking, the detector box is filled with helium gas at atmospheric pressure, 
which reduces beam-gas interactions and unwanted multiple Coulomb scatterings between the target and sensors.

The readout of the sensors was done via $20~\rm cm$ long, copper-based single-layer Flex Print Cables (FPC). 
The non-shielded cables were chosen to minimize 
the material in the acceptance of the TPC, knowing that they may inject pick-up noise into the sensors.


\begin{figure}
\resizebox{\linewidth}{!}{
\includegraphics{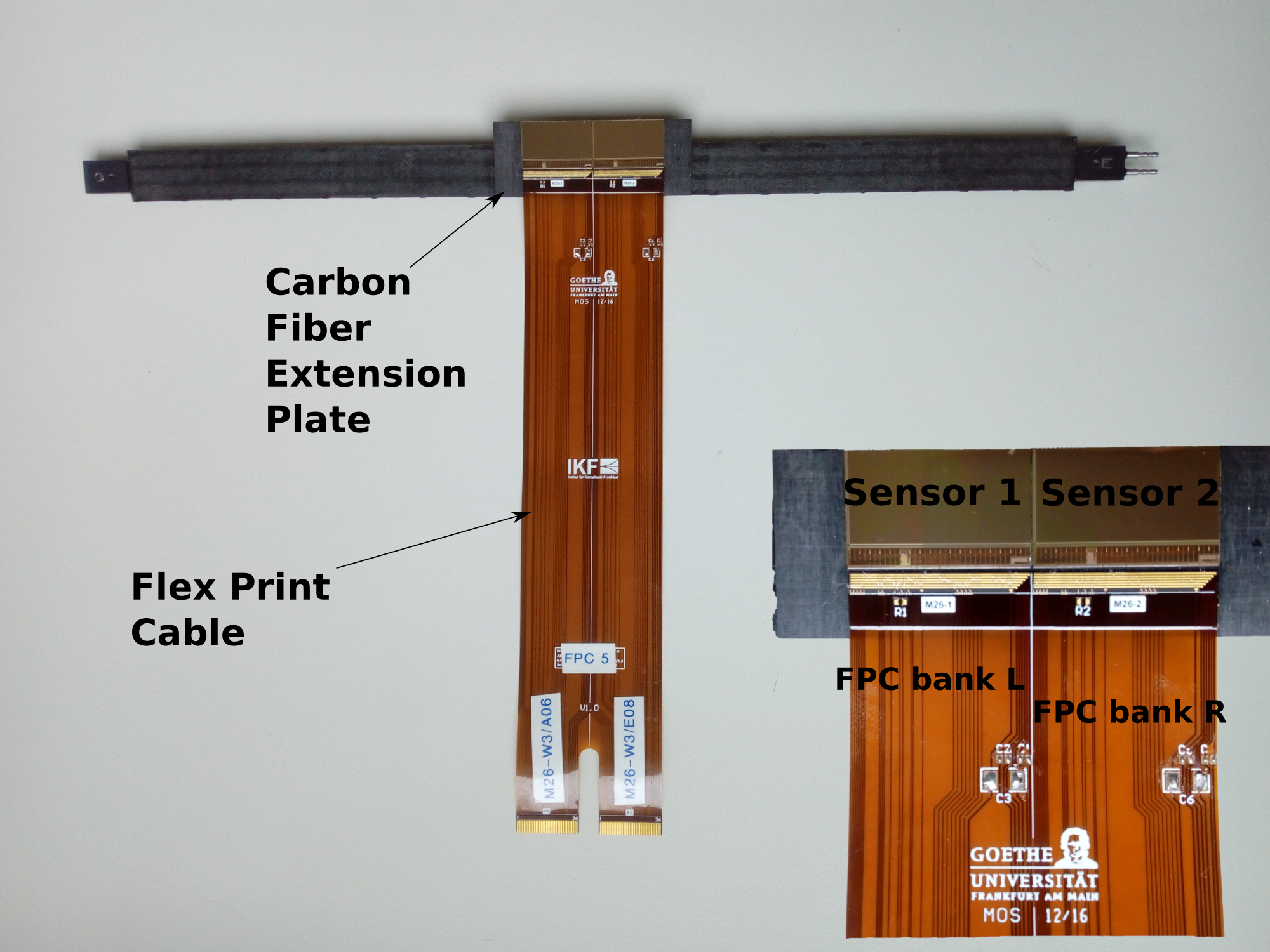}
}
\caption{Single SAVD unit composed of two MIMOSA-26AHR sensors, carbon fibre extension plate, flex-print cable and a supporting ladder. 
The right-bottom part of the figure shows an enlarged view of the sensors, which are also visible in the central part of the ladder.}
\label{fig:VD_unit}
\end{figure}

\subsection{Sensor technology and integration}
\label{sec:3.1}
	
The MIMOSA-26AHR sensors have a 1.06 $\times$ 2.13 cm$^2$ sensitive area, which is covered by 1156 columns 
made of 576 pixels giving 663.5k pixels per chip. 
The pixel pitch is 18.4 $\upmu$m in each direction, which leads to an excellent spatial resolution of 4.5 $\upmu$m. 
The sensor readout is done with a column-parallel 
rolling shutter. The readout time is equivalent to the time resolution of the device and amounts to 115.2 $\upmu$s. 
The slow control of the sensors is done via a JTAG interface, and the most relevant voltages are generated with internal DACs. 
A prominent exception to this rule is the so-called clamping voltage, which has to be provided
from an external source and sets the dark output signal of the pixels. The sensor performs internal signal discrimination, 
zero suppression and 
the first stage of cluster finding. The data is sent out via two 80~Mbps digital links. Four threshold values set for each chip may be set. 
They are shared by the pixels of 289 columns. 

The 50 $\upmu$m thin sensors are flexible and initially slightly bent. Their integration was carried out at the 
Institute of Nuclear Physics (IKF) of the Goethe-University Frankfurt am Main.
The sensors were first glued together with the flex print cable to a base plate made from
carbon fibre. This base plate is used as a mechanical adapter. It is needed as the sensor, and cable size exceeds the ITS ladder's width. 
After gluing, the bending of the sensor was eliminated, and it was wire bonded to the FPC. Finally, the base plate was glued on the 
ladder structure.
A photograph of the module obtained is shown in Fig.~\ref{fig:VD_unit}. The estimated average material budget of the module 
in its active area amounts $\sim 0.3\%~X_0$.

\subsection{The DAQ system of SAVD}
\label{sec:3.2}

\begin{figure*}
\resizebox{\linewidth}{!}{
\includegraphics{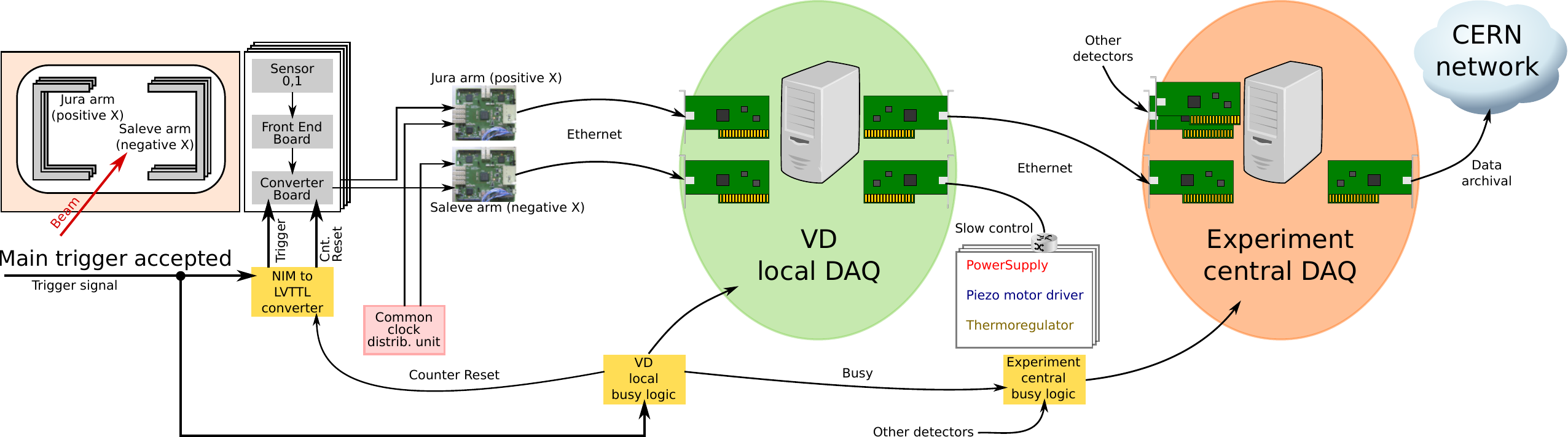}
}
\caption{Schematic diagram of the SAVD readout in the \NASixtyOne experiment.}
\label{fig:VD_Readout}
\end{figure*}

A schematic diagram of the local SAVD DAQ is depicted in Fig.~\ref{fig:VD_Readout}. It relies on hardware and software modules,
which were  initially developed for the prototype of the CBM Micro Vertex Detector \cite{Klaus:2016rty}
and adapted to the needs of SAVD. 

The sensors are connected with the  FPCs to a
Front End Boards (FEB) are located outside of the acceptance on the C-frames. The FEB  boards perform noise filtering.
A conventional flat cable connects the FEBs with the so-called converter boards located at the outer side of the box.
The converter boards host remote-controlled voltage regulators. Moreover, the boards host a latch-up protection system. This 
system monitors the bias currents of the sensors and can detect possible over-currents 
as caused by a 
latch-up. 
If a latch-up
is detected, a rapid power cycle on a given sensor is enforced 
to extinguish the related meta stable short circuit. 

The sensors are steered and read out by two TRBv3 FPGA boards \cite{Paper:TRB3}. The standard TDC firmware of these boards was 
replaced by a dedicated code for steering \linebreak \mbox{MIMOSA-26AHR} sensors. Hereafter, each board serves a readout of eight sensors (data produced in each arm).
During the 2016 test run, the two boards were operated with independent clocks. 
Consequently, the data was synchronized based on the global trigger of \NASixtyOne only. 
Starting from 2017, the boards operated on a common clock, and the sensors remained also synchronized in hardware.

The sensors and the TRBv3 boards operate continuously and stream out their data with the UDP protocol 
through the gigabit-Ethernet interface to a DAQ-PC. 
To synchronize the data with the trigger of \NASixtyOne, the TRBv3 boards receive the trigger signal via the 
converter board. Information on the arrival time of the trigger is added
in real-time to the data stream, but for the sake of simplicity, the data selection is  
performed in software on the DAQ-PC. Five sensor frames per trigger were forwarded to the central
DAQ after the selection was performed, all other data was rejected. The DAQ-PC also performs basic checks on data integrity.  
In the case of inconsistencies suggesting sensor malfunctioning, a sensor reset is scheduled 
and the necessary  reprogramming of the sensors via the JTAG interface is performed during the next spill break.

The central \NASixtyOne DAQ runs in a data push mode.
To prevent mixing events with different trigger numbers,
each subsystem must deliver a busy logic signal.
If any of the detector's busy logic lines are asserted, the whole system 
is halted.
If this waiting time surpasses the delay limit, data acquisition is stopped, and all subsystems run through a restart procedure.
The SAVD busy signal is generated by its local DAQ program using
an external Arduino board.

\section{Detector performance and event reconstruction}
\label{sec:5}

The Vertex Detector was designed for high-efficiency tracking and finding of primary and secondary vertices with high resolution.
The detector concept was developed based on simulations \cite{SAVD_Pb-Pb_add,Ali:2013fea,Ali:2014ewa}. 
The goal was to keep the number of sensors low while requiring  the system covers most of the produced open charm mesons.

\begin{figure}[h]
  \resizebox{\linewidth}{!}{
		\includegraphics{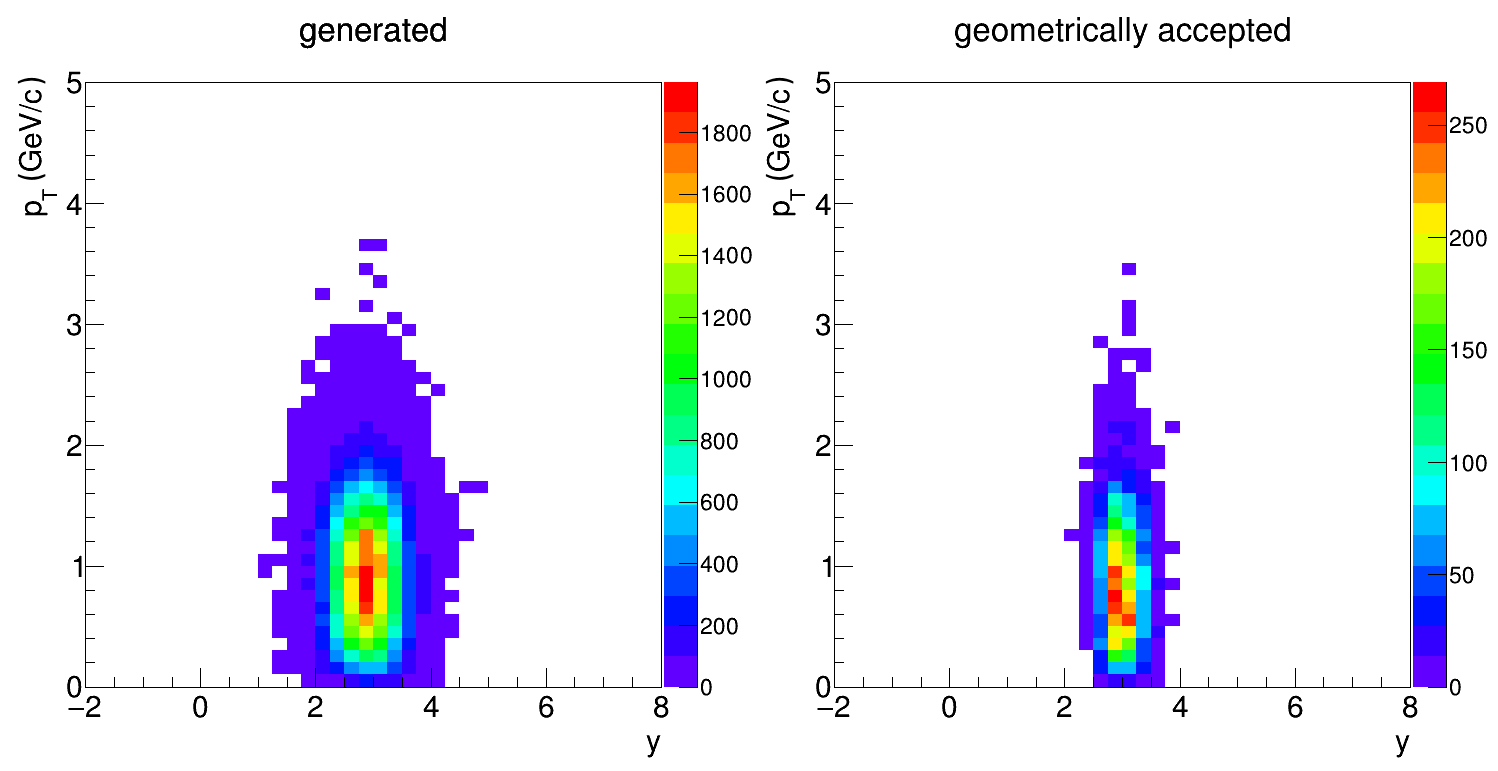}
    }
  \resizebox{\linewidth}{!}{
		\includegraphics{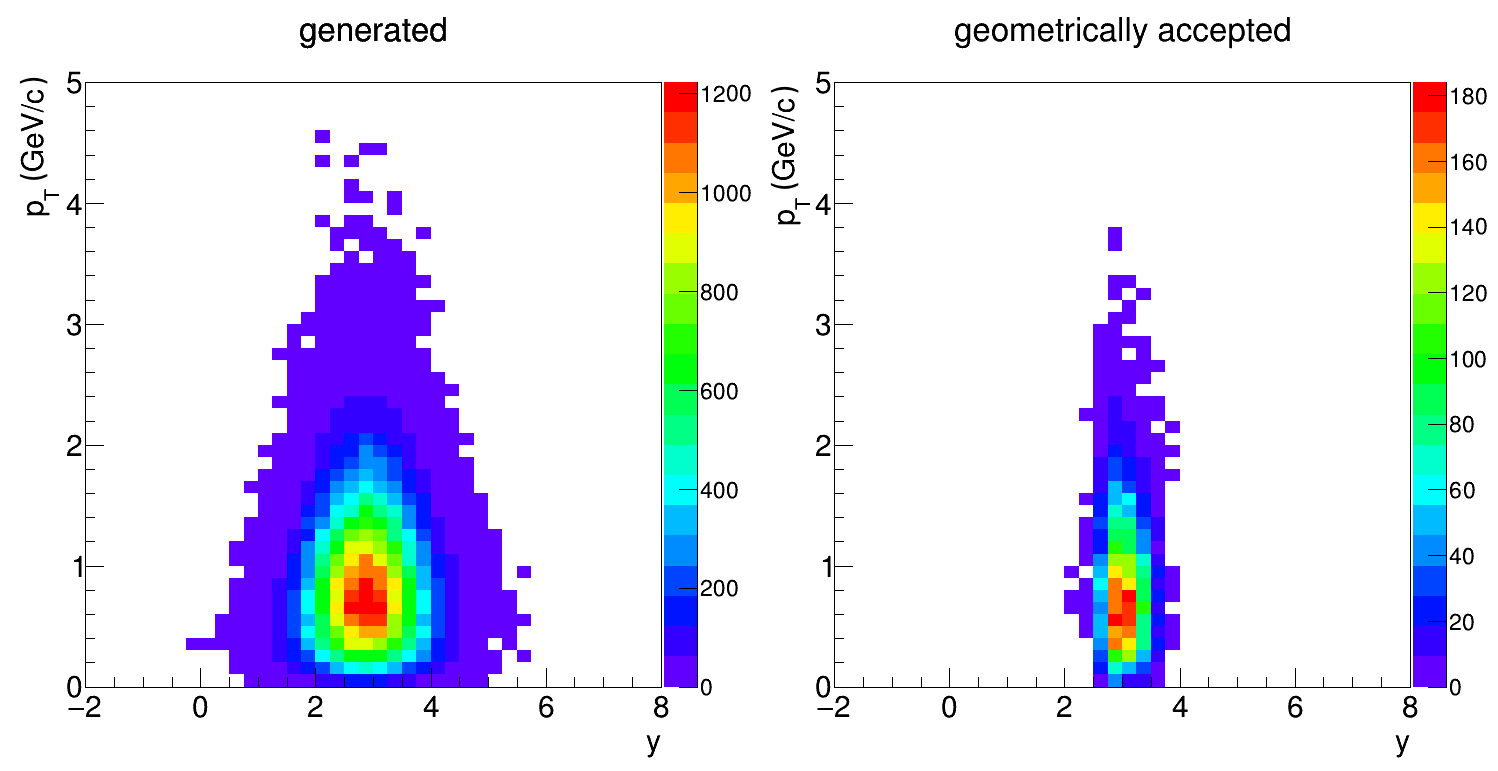}
    }
\caption{Rapidity -- transverse momentum spectra of $D^0$ + $\overline{D^0}$ mesons for 20\% of the most central Xe+La  collisions at  150\AGeVc according to AMPT (top plots) and PHSD (bottom plots) predictions. 
          The left plots show the generated phase space, and the right plots show geometrically accepted phase space. 
          The plots are obtained for 1M generated $D^0$ + $\overline{D^0}$ decaying in the $\pi$ and $K$ channel.}
	\label{fig:SAVDacceptance_xela}
\end{figure}

For studying the detector efficiency and acceptance, the simulations were performed using the \GeantFour package (for more details, see Ref.~\cite{Merzlaya:2021kue}).
The background was described using the AMPT model~\cite{Lin:2004en} and for the \linebreak parametrization of 
the open charm meson spectra, the AMPT and the PHSD models were used. 
Figure~\ref{fig:SAVDacceptance_xela} presents the distribution 
of transverse momentum -- rapidity of all generated $D^0$ and $\overline{D^0}$ and 
 those  $D^0$ and $\overline{D^0}$, that pass the detector acceptance, i.e. when both of the daughter tracks have sufficient for reconstruction number of SAVD (3 or more) and TPC (10 or more) hits.
The simulation  for Xe+La collisions at 150\AGeVc shows that about 7.8~\% 
and 5.9~\% acceptance  of $D^0 + \overline{D^0}$ in $\pi$ and $K$ decay channel  for AMPT and PHSD phase space distributions, respectively.

\subsection{Sensor operation and efficiencies}
\label{sec:sensor-operation}

In SAVD, the sensors are located as close as 3~mm from the beam center. Thus
they are exposed to primary beam ions from the beam halo and  
nuclear beam fragments. It was considered that the related impacts would
create latch-up and do severe damage
to the sensors. Fortunately, although the beam halo ranged
to 1~cm from the beam axis, this was not the case. The ion impacts were observed to create clusters. 
of the size up to 200 pixels, but no sensor 
was destroyed by the radiation   
during the detector operation. This was certainly a success of the related protection 
system and reflected the unexpectedly good robustness of the sensors.  

A dedicated radiation test has shown that 30\AGeVc Pb ions
created an integrated, non-ionizing radiation damage of $\lesssim 300~\rm n_{eq}/cm^2$ (upper limit).
As expected by our radiation dose estimates, the radiation damage in the sensors
remained below the radiation tolerance of the sensor, which amounts to $\sim 150~\rm~krad$
and $\gtrsim 10^{13}~ \rm n_{eq}/cm^2$ at modest cooling (typically the coolant temperature
was chosen with $10~^\circ \rm C$). 

Due to a lack of resources, no near-time monitoring providing a sensor detection efficiency was available
during the data taking. The thresholds of the sensors were thus lowered until the highest reasonable dark occupancy 
of $\sim 10^{-4}$ was reached. Based on the sensor's known efficiency/dark occupancy curve, we expected
to reach a good efficiency. However, disappointing efficiencies of $10-94~\%$ were observed in the 2016 Pb test run, and two sensors did not work.
This lack of efficiency was dominantly caused by a bad synchronization of the data selected by the trigger, which
rejected valid data in some cases. This was corrected for the 2017 Xe-La run. Moreover, the biasing voltages 
were adapted for the nominal settings to account for the ohmic losses
in the FPCs. Still, the impact on the clamping voltage had not been considered properly. This issue generated
a saturation of the pre-amplifiers of multiple pixels. Once identified, it was corrected by adapting 
a reference voltage of the pre-amplifiers by slow control.

Thanks to the modifications and sensor repairs, all sensors were operational in the 2017 Xe+La run.  
Unfortunately, the above-mentioned coarse approach for threshold tuning had to be used again.
Still, an efficiency between $84~\%$ and the nominal $>99~\%$ was observed and most 
sensors showed an efficiency significantly above $90~\%$. 

\subsection{SAVD internal geometry calibration}
\label{sec:4.1}

\begin{figure}
\resizebox{\linewidth}{!}{
\includegraphics{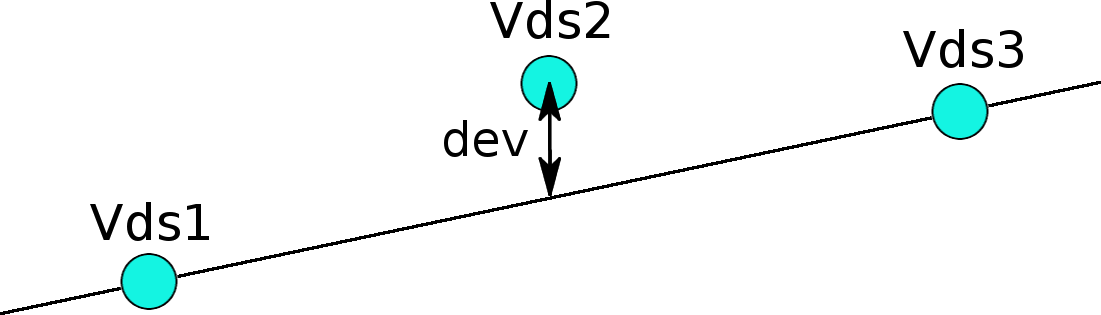}
}
\caption{The graphical representation of the ``dev'' variable used for geometry tuning.}
\label{fig:devVariable}
\end{figure}

\begin{figure}
\centering
\resizebox{.9\linewidth}{!}{
\includegraphics{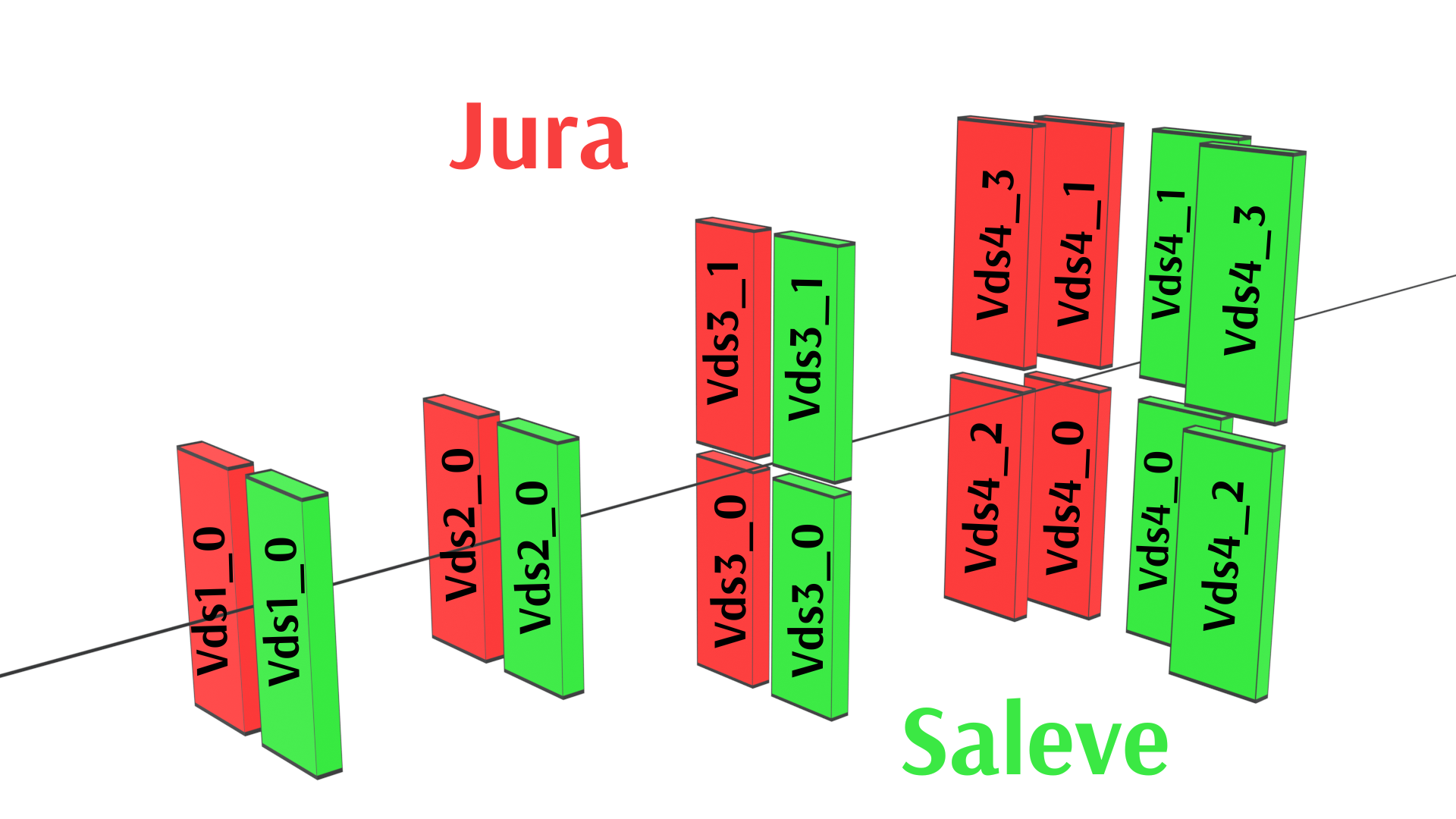}
}
\caption{The naming convention of the SAVD sensors. The first number following ``Vds'' denotes the station, while the second gives the sensor number in a given station.}
\label{fig:sensorsNames}
\end{figure}

The alignment of SAVD was done using track candidates found by the combinatorial method with data 
taken with zero magnetic fields.
The purpose of geometry tuning is to find the corrections for the sensor positions (each sensor has 6 degrees 
of freedom: offsets from the nominal geometry in \coordinate{x}, \coordinate{y} and \coordinate{z} position and 
rotation along \coordinate{x}, \coordinate{y} and \coordinate{z} axes).
For correct geometry alignment, hits produced by the same particle should lie in a straight line.
To define the collinearity of three hits, the variable ``dev'', which represents the 
deviation of the position of the middle cluster from the straight line connecting the other two clusters,  was introduced:

\begin{equation}
\begin{split}
\text{dev}_\mcoor{x} \;=\; \frac{(\mcoor{z}_3-\mcoor{z}_2)\,\mcoor{x}_1\;+\;(\mcoor{z}_2-\mcoor{z}_1)\,\mcoor{x}_3}{\mcoor{z}_3-\mcoor{z}_1}\;-\;\mcoor{x}_2,
\\
\text{dev}_\mcoor{y} \;=\; \frac{(\mcoor{z}_3-\mcoor{z}_2)\,\mcoor{y}_1\;+\;(\mcoor{z}_2-\mcoor{z}_1)\,\mcoor{y}_3}{\mcoor{z}_3-\mcoor{z}_1}\;-\;\mcoor{y}_2,
\end{split}
\end{equation}

where the variables are explained in Fig.~\ref{fig:devVariable}.
For properly calibrated internal geometry, the distribution of the ``dev'' variables should show a narrow correlation peak centered at zero. The positions resolutions in \coordinate{x} and \coordinate{y} directions can be then determined from the obtained distributions, which are approximately
equal to $\sqrt{\frac{2}{3}}\;\sigma_{\mcoor{x},\mcoor{y}}$, were  $\sigma_{\mcoor{x}}$($\sigma_{\mcoor{y}}$) represent the width of the $\text{dev}_\mcoor{x}(\text{dev}_\mcoor{y})$ distribution. The factor $\frac{2}{3}$ refers to the equal Vds1 to Vds2 and Vds2 to Vds3 distances in \coordinate{z} coordinate (see Fig.~\ref{fig:devVariable}).


The calibration  algorithm uses the MIGRAD function of the MINUIT~\cite{James:2004xla} package. 
The Variable Metric method was used to minimize the ``dev'' function
to find the optimal alignment parameters.

\begin{figure}
\resizebox{1.05\linewidth}{!}{
\includegraphics{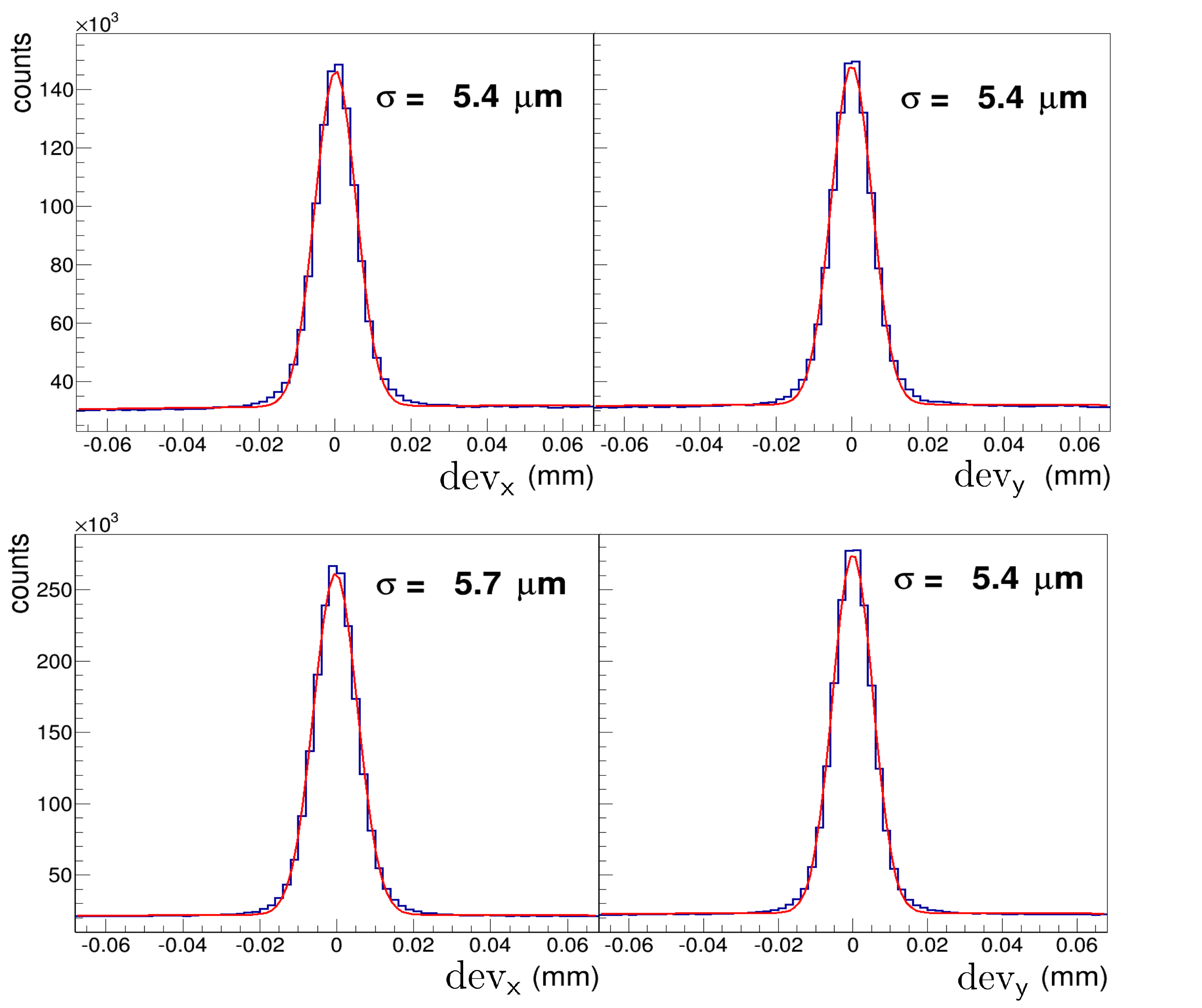}
}
\caption{Distribution of ``dev'' in \coordinate{x} (\textit{Left}) and \coordinate{y} (\textit{Right}) coordinate for
two different sensors combinations: Vds1\_0, Vds2\_0, Vds3\_1 (\textit{Top}) and  Vds2\_0, Vds3\_0, Vds4\_0 (\textit{Bottom}).
The red lines represent fits with the sum of the Gaussian function (signal component) and the second-order 
polynomial (combinatorial background).}
\label{fig:devExample}
\end{figure}
 
A detailed description of the applied geometry reconstruction procedure is provided \cite{Brylinski:2633136}.
It is seen from the plot presented in Fig.~\ref{fig:devExample}, that the obtained position resolution provided by sensors 
is on the level of the nominal 4.5~$\mu$m in both \coordinate{x} and \coordinate{y} coordinates.

\subsection{Cluster, track and vertex reconstruction}
\label{sec:5.1}

	
The first step of data reconstruction is cluster recognition. 
A particle passing through a sensor may fire more than one 
pixel in a given sensor.
These pixels should thus not be considered to indicate independent particle hits but rather together constituting a particle hit.
Such a composite particle hit is called a ``cluster''.
A computer algorithm, the so-called ``clusteriser'', identifies such clusters.
It takes each pixel as a starting point and searches neighboring pixels containing signals in both dimensions.
The search is repeated recursively for neighboring fired pixels until no more neighboring fired pixels can be found.
The set of fired pixels is used to calculate the center of gravity, taken as the center of the resulting cluster.


The tracks registered in SAVD are
slightly curved because of the magnetic field.
This curvature is small enough to use a straight line to identify clusters in different stations on the same track.
Consequently, a straight line was chosen 
to describe the tracks:
\begin{equation}
\begin{split}
&\mcoor{x}(\mcoor{z})=A\mcoor{z}+\mcoor{x}_0,\\
&\mcoor{y}(\mcoor{z})=B\mcoor{z}+\mcoor{y}_0.
\end{split}
\end{equation}
Using this parametrization, a combinatorial track identification procedure based on 
checking the combinations of all hits from different stations was introduced.  
If the hits detected on different SAVD stations lie on a straight line according to a $\chi^2$ criterion, 
the combination is accepted as a reconstructed track.

The track reconstruction procedure was first implemented for the field-off data set. 
From the distributions of the residuals of hits from the reconstructed and fitted with a straight 
line tracks, the spatial sensor resolution was determined to be on the level of 5~$\upmu$m, as was expected.

It turned out that the same straight-line combinatorial method could also be applied to reconstruct the tracks for physics data sets with the field on.
However, if the straight-line track model is applied, the hits on the third and fourth stations of SAVD visibly deviate from the fitted straight line. 
The result of this is a double-peak structure in the distribution of cluster deviations for the x-direction 
rather than a Gaussian distribution. 
This effect is caused by the vertical $B_y$ component of the magnetic field in the SAVD volume. 
Therefore in the next steps of the reconstruction, the positions of hits are fitted using a 
second-order polynomial function for \coordinate{x} and linear for \coordinate{y} coordinate:
\begin{equation}
\begin{split}
&\mcoor{x}(\mcoor{z}) = A_2 \, \mcoor{z}^2+A_1\mcoor{z}+\mcoor{x}_0,\\
&\mcoor{y}(\mcoor{z}) = B \mcoor{z}+\mcoor{y}_0.
\end{split}
\end{equation}

\begin{figure}
\resizebox{\linewidth}{!}{
\includegraphics{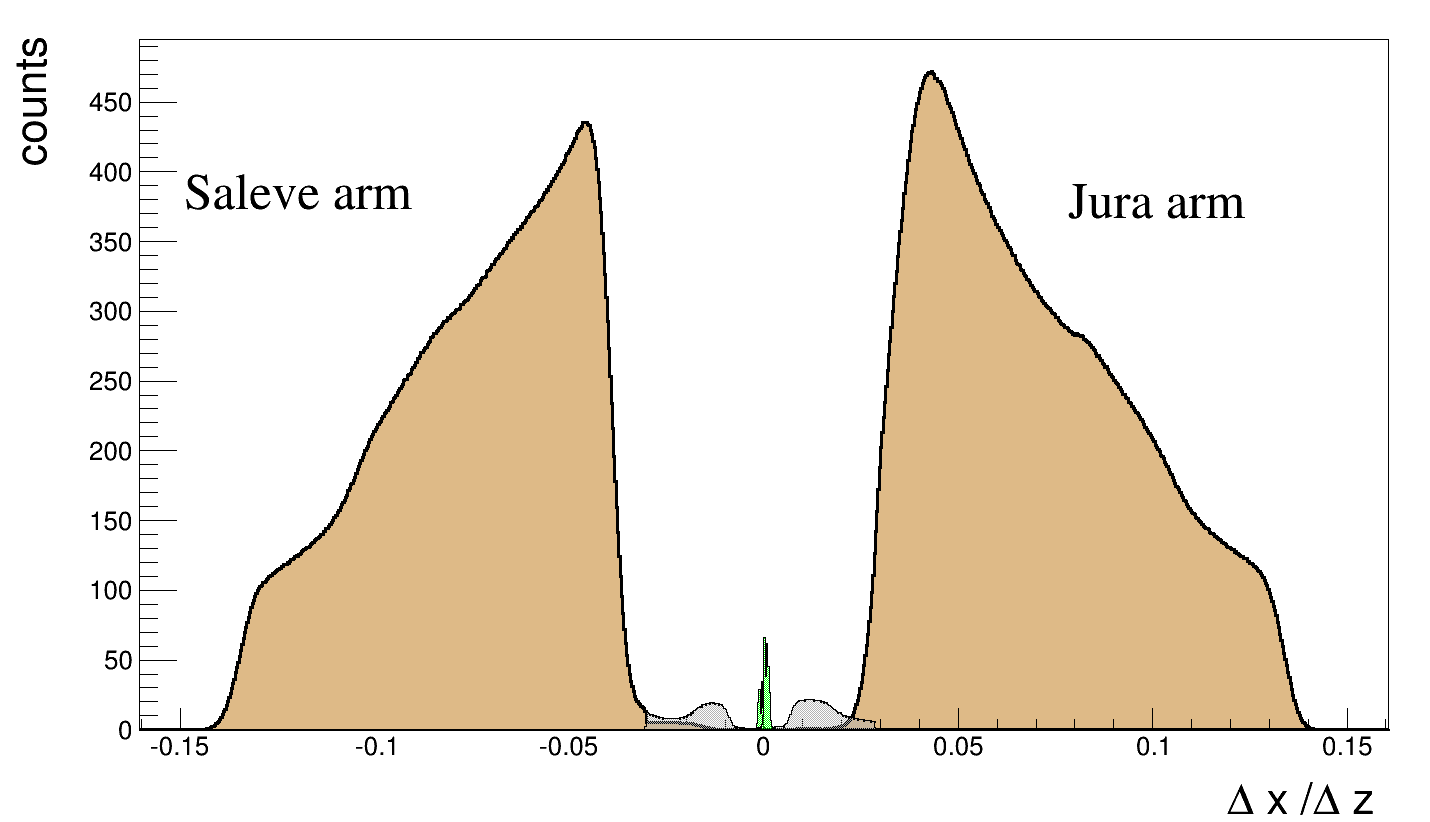}
}
\caption{
$\Delta \mcoor{x}/\Delta \mcoor{z}$ (\coordinate{x}-slope) distribution tracks reconstructed in Jura (positive values) and Saleve (negative values) arms. The plot is done for Xe+La at 150\AGeVc data set taken in 2017. Different colors refer to different production components explained in the text.
}
\label{fig:ax-components}
\end{figure}

The distribution of the ${\Delta \mcoor{x}}/{\Delta \mcoor{z}}$ ratios for the reconstructed tracks is shown in Fig.~\ref{fig:ax-components}. 
The ratios are calculated for track lines reconstructed in the target region, referring to tracks emission angles in the \coordinate{xz}-plane. 
The distribution reflects a clear three-peak structure for each arm. 
Firstly, the narrow inner-most peak (green peak at small angles) is associated with particles produced far upstream and traveling 
parallel to the beam for a long distance. 
Next, the middle structure (gray histogram) corresponds to particles produced upstream of the target. 
Finally, the outer peak (brown color histogram) is generated by particles produced in the target - these tracks are selected for further analysis. 
	
The primary vertex is the point of the closest convergence of all reconstructed tracks.
Thus, the longitudinal coordinate of the primary vertex is found by minimizing the expression:
\begin{equation}
\begin{gathered}
D(\mcoor{z})=  \\
\sum_{i<j}\lbrace(A_i\mcoor{z}+\mcoor{x}_i^0-A_j\mcoor{z}-\mcoor{x}_j^0)^2+(B_i\mcoor{z}+\mcoor{y}_i^0-B_j\mcoor{z}-\mcoor{y}_j^0)^2 \rbrace,
\end{gathered}
\end{equation}
which describes the sum of 
squares of
the relative distances of all track pairs reconstructed in a single event at the given transverse plane 
defined by the longitudinal coordinate \coordinate{z}.
The $\mcoor{x}_{prim}$ and $\mcoor{y}_{prim}$ coordinates of the primary vertex are afterwards 
calculated as the average of \coordinate{x} and \coordinate{y} positions of tracks at $\mcoor{z} = \mcoor{z}_{prim}$. 
	
\begin{figure}
\resizebox{\linewidth}{!}{
\includegraphics{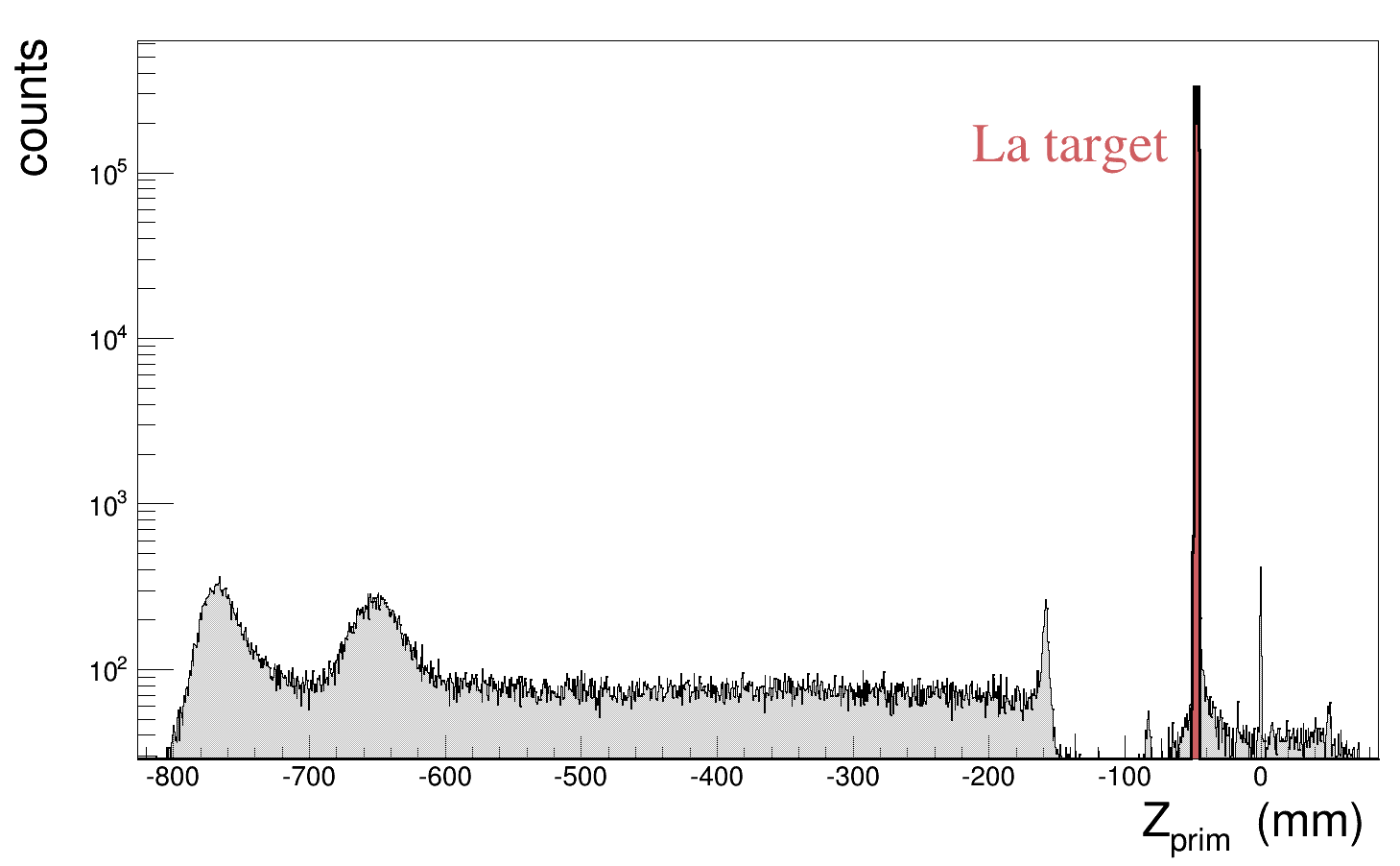}
}
\caption{
Distribution $\mcoor{z}_{prim}$ of primary vertices for tracks produced on target (brown color histogram) 
and production out of the target (dark color histogram). Seed text for more explanation. 
}
\label{fig:VtxZ_components}
\end{figure}

To support the interpretation of components from Fig. \ref{fig:ax-components}, the primary vertex reconstruction 
was performed on the event by event basis separately for 
tracks within the $|\Delta \mcoor{x}/\Delta \mcoor{z}|$
the interval from 0.01 to 0.025 (gray histogram) and with  $|\Delta \mcoor{x}/\Delta \mcoor{z}| > 0.025$ (brown histogram). 
By looking at the longitudinal distribution of the primary vertex for these samples of tracks 
(see Fig.~\ref{fig:VtxZ_components}) it can be seen that, indeed, the tracks associated with the
the most outer peak in Fig.~\ref{fig:ax-components} (brown) originates from the target, which is located 47~mm upstream 
from the first VD station.
The primary vertices associated with tracks from the middle peak (gray in Fig.~\ref{fig:ax-components}) 
are relatively smoothly distributed upstream of the target in the range from -1200 mm 
(exit from the beam-line) to -50~mm (near the target).
At -190~mm, the distribution has a sharp peak related to interactions in the aluminized Mylar front window of the SAVD box.
One can also see that between the window and the target, the frequency of interaction drops by a factor of 5 due to 
the presence of helium gas in the SAVD vessel.



\begin{figure}
\resizebox{\linewidth}{!}{
\includegraphics{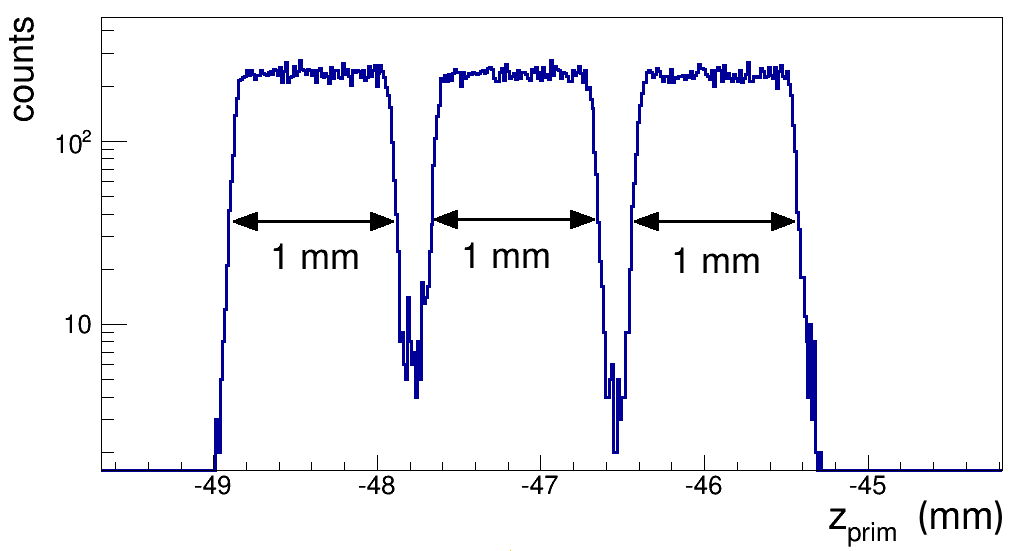}
}
\caption{Distribution of of the longitudinal coordinate $\mcoor{z}_{prim}$ for the Xe+La data at 150 \AGeVc recorded in 2017.}
\label{fig:primaryZ_xela}
\end{figure}

A target segmented to three 1~mm thick La layers was used for Xe+La data taking.
The target structure can be well seen in the $\mcoor{z}_{prim}$ distribution shown in Fig.~\ref{fig:primaryZ_xela}.
It is seen that the precision of the primary vertex reconstruction allows for determining on which particular
target segment the Xe+La collision occurred.  

\begin{figure}
\resizebox{\linewidth}{!}{
\includegraphics{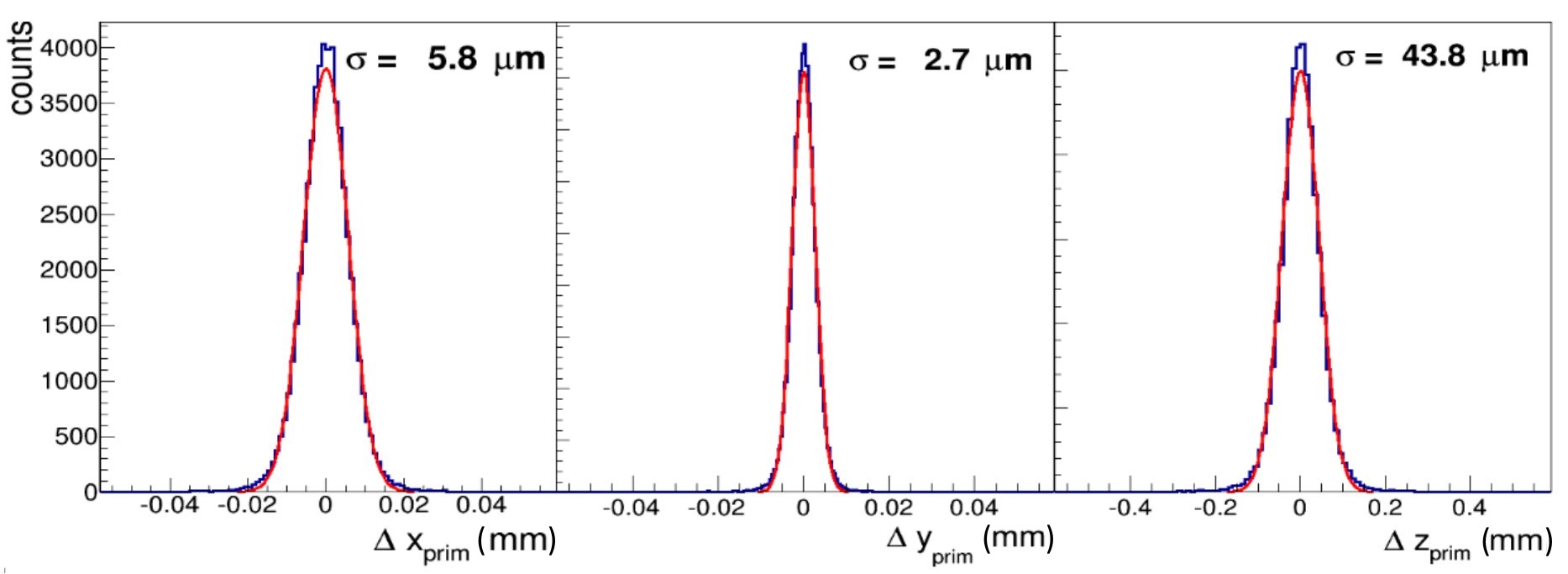}
}
\caption{Distributions of differences between \coordinate{x}, \coordinate{y} and 
\coordinate{z} coordinates of the primary vertexes 
reconstructed using sub-event~1 and sub-event~2 tracks (see text for details) 
for the Xe+La data at 150\AGeVc recorded in 2017.}
\label{fig:primary_vert_res}
\end{figure}

To determine the spatial resolution of the primary vertex reconstruction, the SAVD tracks from an event were split into two
non-overlapping sub-events, namely every second track from Jura and Saleve arms, were assigned to sub-event~1, whereas 
the remaining tracks were assigned to sub-event~2. 
In this way, one obtains two equivalent track samples. The primary vertex spatial resolutions obtained with 
sub-event~1 and sub-event~2 are expected to be identical since the opening angle range for both samples is the same. 
The distributions of differences between \coordinate{x}, \coordinate{y} and \coordinate{z} coordinates 
of the primary vertices reconstructed using 
sub-event~1 and sub-event~2 tracks are shown in Fig.~\ref{fig:primary_vert_res} 
for the Xe+La data.
The red lines correspond to Gaussian fits of the distributions.
The observed widths of the peaks can be converted to the spatial resolution of the primary 
vertex, namely $\sigma_\mcoor{x}$ = 1.3~$\upmu$m, $\sigma_\mcoor{y}$ = 1~$\upmu$m and $\sigma_\mcoor{z}$ = 15~$\upmu$m, 
for \coordinate{x},  \coordinate{y} and \coordinate{z} coordinate, respectively.

After the primary vertex is found, the next step of track reconstruction searches for 
tracks using the Hough transform (HT) method (for details, see Ref.~\cite{Merzlaya:2017nvb}).
It is a global method of track reconstruction where each cluster is processed only once.
Thus, the computation time of this method is proportional to the number of all detected hits and is much faster 
than the combinatorial method, 
which accesses clusters in the nested loops over clusters grouped according to the station of their detection.
However, the HT method requires information about the origin point
thus, it is implemented as a second step of the SAVD track reconstruction chain.
The HT procedure is based on representing the track as a set of two 
slope parameters $\left( a_\mcoor{x}, a_\mcoor{y}\right)$, 
which can be used to describe straight track lines according to the following parametrization:
\begin{equation}
\begin{split}
&\mcoor{x}(\mcoor{z})=a_\mcoor{x} \mcoor{z},\\
&\mcoor{y}(\mcoor{z})=a_\mcoor{y} \mcoor{z},
\end{split}
\end{equation}
where $\mcoor{x}, \mcoor{y}, \mcoor{z}$ are cluster coordinates with respect to the primary vertex position.
Then, for each hit its position in coordinate space $\left( \mcoor{x}, \mcoor{y} ,\mcoor{z}\right)$ are 
transformed to so-called Hough space of parameters $\left( a_\mcoor{x}, a_\mcoor{y}\right)$. 
Further, hits left by the same particle would have the same track parameters and appear as peaks in the Hough 
space presented as a 2-dimensional histogram. The algorithm searches for such local peaks which correspond 
to tracks. However, due to multiple scattering and track curvature, hits that belong to the same track might 
appear in different bins of the Hough space histogram. Thus, the algorithm performs the clusterisation procedure: 
combining neighboring bins into one cluster.

	

\subsection{Track reconstruction efficiency}

\begin{figure}[h]
\resizebox{\linewidth}{!}{	
\includegraphics{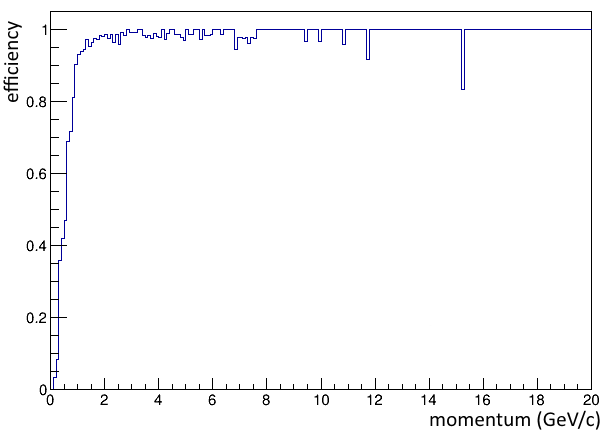}
}
	\caption{The SAVD reconstruction efficiency versus track momenta assuming fully efficient 
          sensors for Xe+La at 150\AGeVc.}
\label{fig:savd_pureeff}
\end{figure}

To test the reconstruction efficiency of SAVD, the \GeantFour-based simulation study was performed (the effect of the sensor inefficiency was excluded). 
The efficiency was determined as the ratio between the number of the reconstructed 
SAVD tracks and the number of the simulated SAVD tracks with three and four hits.
Fig.~\ref{fig:savd_pureeff} shows the dependence of the efficiency versus track momenta. It is seen that the efficiency 
is close to 100~\% for higher track multiplicity. However, it 
starts to drop
for tracks with momentum $<$~1~GeV/$c$.

Low momentum tracks have large curvature in the SAVD region (the magnetic field is low but not zero). 
Thus such tracks 
can neither
be reconstructed within the straight-line model of the combinatorial reconstruction, 
nor during the Hough Transform stage 
as the hits belonging to these tracks 
are transformed into the different Hough space regions. 

\subsection{VD--TPC global geometry calibration}
\label{sec4.2}
\begin{figure}
\resizebox{\linewidth}{!}{
\includegraphics{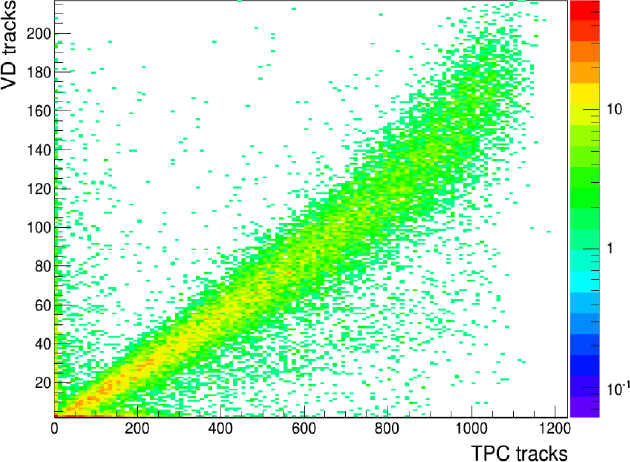}
}
\caption{TPC track multiplicity versus SAVD track multiplicity; for detail, see text.}
\label{fig:Figure12}
\end{figure}

The track multiplicity correlation between tracks reconstructed (all collected events, no trigger selection) in  SAVD and  TPCs is shown in Fig.~\ref{fig:Figure12}. 
As one can see, the multiplicities of SAVD and TPC tracks are well correlated, proving that the tracking procedures described above are correct.
It may be observed that for some events, tracks were reconstructed in either VD or TPCs, but not both. These cases are related to low-track multiplicity events selected by the minimum bias trigger.

Merging the track fragments measured by SAVD and TPCs requires the SAVD alignment relative to the TPCs. 
By observing the difference between the positions of reconstructed primary vertices in the SAVD and 
the TPCs in a given event, the SAVD position was calibrated with an accuracy 
of 16~$\upmu$m, 6~$\upmu$m and 100~$\upmu$m in the \coordinate{x}, \coordinate{y} and \coordinate{z} 
coordinate, respectively.

\subsection{Global tracking}
\label{sec:5.2}

\begin{figure}
\resizebox{\linewidth}{!}{
\includegraphics{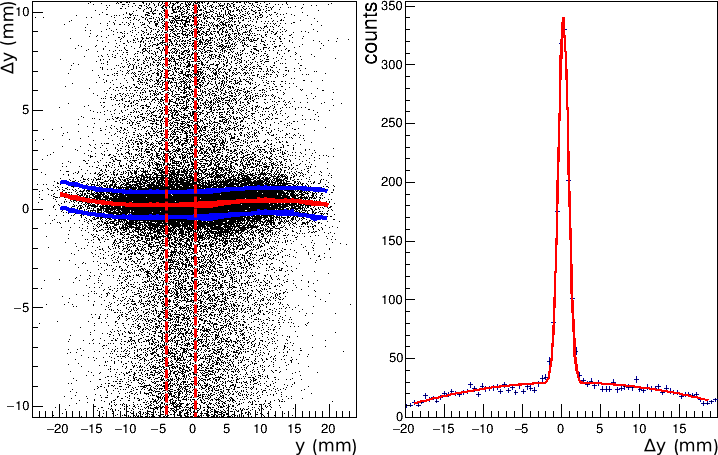}
}
\caption{
\textit{Left}: Difference in $\mcoor{y}$ coordinate of SAVD and TPC tracks ($\Delta \mcoor{y}$) versus $\mcoor{y}$ 
at the merging plane.
\textit{Right}: example of the projection of distribution of $\Delta \mcoor{y}$ versus $\mcoor{y}$ 
onto the $\Delta \mcoor{y}$  
coordinate for $-7$~mm $< \mcoor{y} < -2.5$~mm (single slice).
}
\label{fig:matching}
\end{figure}

The merging of SAVD and TPC track fragments is done in three steps:
\begin{enumerate}[(i)]

\item Since tracks are not affected by the magnetic field in the $\mcoor{y}$ direction, all SAVD tracks 
are combined with VTPC tracks, and for each SAVD--VTPC track pair, the difference between the tracks 
slopes in the $\mcoor{y}$ coordinate, $\Delta a_\mcoor{y}$, is calculated.
The distribution of $\Delta a_\mcoor{y}$ shows a sharp peak on a large combinatorial background.
A $\pm \, 5 \sigma$ cut around this peak is applied to pre-select SAVD and TPC track pairs that potentially match.

\item For a given track pair, the TPC momentum is assigned to the SAVD track.
This allows extrapolating the SAVD track to the VTPC front surface where both are 
matched in $\mcoor{x}$, $\mcoor{y}$ ($\mcoor{z}$ is matched by construction as it defines the merging plane) 
coordinates and the difference of the track positions $\Delta \mcoor{x}$ and $\Delta \mcoor{y}$ are calculated.
Fig.~\ref{fig:matching}~\textit{(left)} shows the distribution of $\Delta \mcoor{y}$ versus $\mcoor{y}$ 
of \textit{Saleve} side SAVD tracks matched to \textit{Jura} side tracks of VTPC1.
Because the average value of $\Delta \mcoor{y}$ depends on $\mcoor{y}$, narrow ranges of $\mcoor{y}$ of the 
distribution are projected onto $\Delta \mcoor{y}$.
The projected distributions (slices) are then fitted with a sum of a second-order polynomial which describes 
the background related to false-merging cases and a Gaussian peak that accounts for the true ones.
An example of a single slice is shown in Fig.~\ref{fig:matching} \textit{(right)}.
The dependence of the fitted mean ($\left< \Delta \mcoor{y} \right>$) and variance ($\sigma_{\Delta \mcoor{y}}$) 
on $\mcoor{y}$ are then 
fitted with a third-order polynomial function. 
The results of these fits are shown as red ($\left< \Delta \mcoor{y} \right>(\mcoor{y})$) and blue lines 
($\pm$ $\sigma_{\Delta \mcoor{y}}(\mcoor{y})$) in Fig.~\ref{fig:matching}~\textit{(left)}. 
A similar procedure was used for $\Delta \mcoor{x}$ versus $\mcoor{z}$ merging.
Both $\Delta \mcoor{y}$ versus $\mcoor{y}$ and $\Delta \mcoor{x}$ versus $\mcoor{z}$ distributions were constructed 
for \textit{Jura} - \textit{Jura}, \textit{Jura} - \textit{Saleve}, 
\textit{Saleve} - \textit{Saleve} and \textit{Saleve} - \textit{Jura} track combinations, 
separately for VTPC1 and VPTC2.

\item The values of $\left< \Delta \mcoor{y} \right>$, $\sigma_{\Delta \mcoor{y}}$ and $\left<\Delta \mcoor{x}\right>$, 
$\sigma_{\Delta \mcoor{x}}$ obtained from the fits are used to apply elliptic cuts to select the best merge candidate.  

\end{enumerate}

\begin{figure}
\resizebox{\linewidth}{!}{
\includegraphics{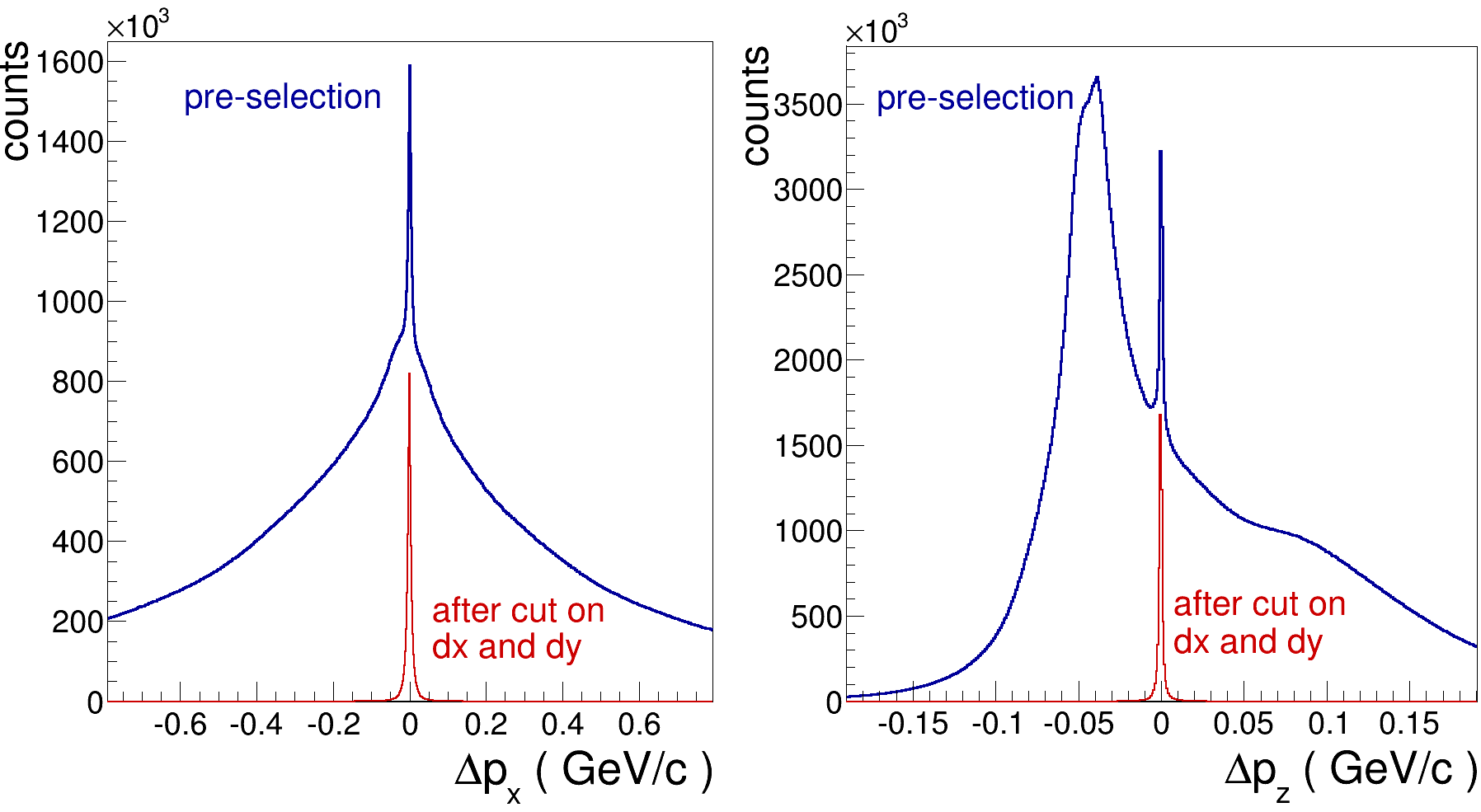}
}
\caption{Difference of momentum components $\Delta p_\mcoor{x}$ and $\Delta p_\mcoor{z}$
calculated at the merging plane for SAVD - TPC track combinations that passed the cut on $\Delta a_\mcoor{y}$ 
(blue) and after additional elliptical 4$\sigma$ cuts on $\Delta \mcoor{x}$ and $\Delta \mcoor{y}$ (red).}
\label{fig:matching:pxpz}
\end{figure}

Fig.~\ref{fig:matching:pxpz} shows the distribution of the difference between SAVD and TPC momentum 
components $\Delta p_\mcoor{x}$ and $\Delta p_\mcoor{z}$ calculated at the merging plane 
for SAVD and TPC track combinations that passed the cut on $\Delta a_\mcoor{y}$ (blue) and with the 
elliptical 4$\sigma$ cuts on $\Delta \mcoor{x}$ and $\Delta \mcoor{y}$ (red).
It can be seen that after the $\Delta \mcoor{x}$ and $\Delta \mcoor{y}$ cuts, the distributions 
are practically free of background.

About 75~\% of the SAVD tracks are merged with the VTPC tracks.
This result corresponds to the performed \GeantFour-based simulations.
The remaining tracks either miss the VTPC acceptance, decay before reaching the VTPC or are not merged 
due to the SAVD-TPC merging inefficiency, which is about 5~\%.

Finally, the global track, which has hits in both SAVD and TPCs, is refitted using a method 
based on Kalman Filter \cite{Gorbunov:kalman-filter} and used for further analysis.

\subsection{Secondary vertex resolution}

\begin{figure}[h]
\resizebox{\linewidth}{!}{
	\includegraphics{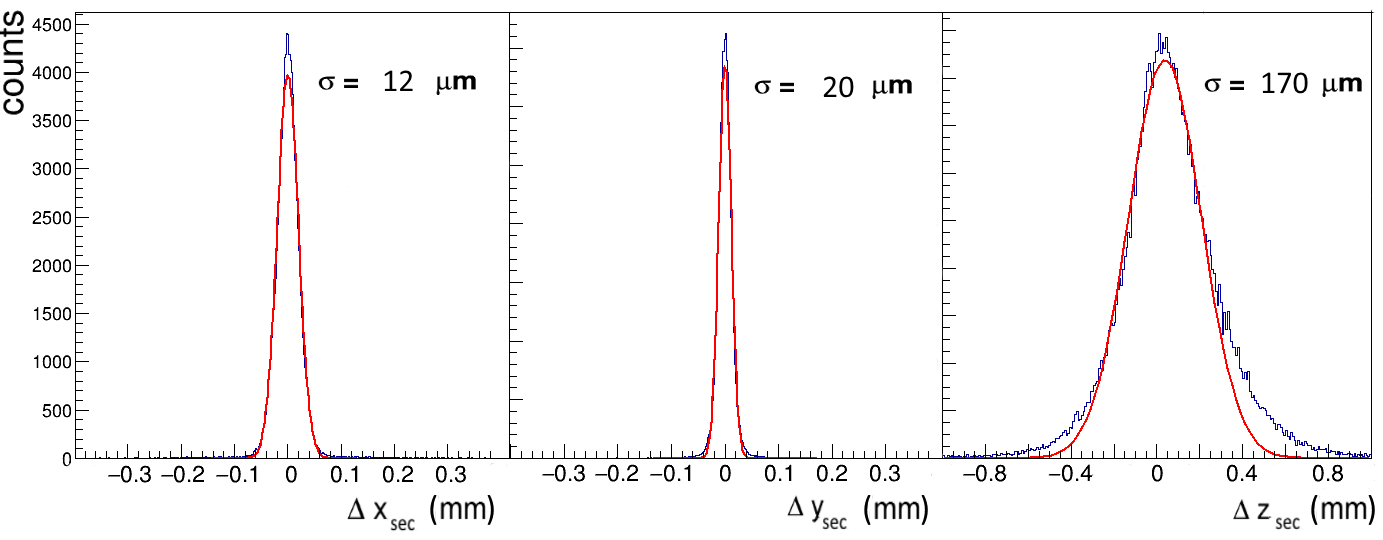}
	}
	\caption{Distributions of differences between coordinates of the reconstructed secondary 
          vertices and simulated vertices for Xe+La at 150\AGeVc data. Lines correspond to Gaussian 
          fits of the distributions, and the $\sigma$ parameters resulting from the fits are indicated.}
	\label{fig:secvertexres_simrec}
\end{figure}

The position resolution of the reconstructed secondary vertices related to open charm mesons decays 
was determined in the \GeantFour-based simulations by comparing the simulated and reconstructed positions of 
the vertices. The differences $\Delta\mcoor{x}$, $\Delta\mcoor{y}$, $\Delta\mcoor{z}$ of the coordinates 
of the reconstructed secondary vertex 
position and the one defined in the \GeantFour-based simulations are shown in Fig.~\ref{fig:secvertexres_simrec}. 
The sigma of these distributions determines the primary-vertex resolution 
to be 20~$\mu$m, 11~$\mu$m, 170~$\mu$m for  $\mcoor{x}$, $\mcoor{y}$ and $\mcoor{z}$ coordinates, respectively. 

\subsection{Invariant mass spectra in Xe+La data}
\label{sec:5.3}

The performance results are based on the 2017 Xe+La data since it is currently the most thoroughly investigated data set.
The SAVD tracks matched to TPC tracks are used to search for the $D^0 + \overline{D^0}$ signal. 
The particle identification (PID) information was not used in the analysis.
Each SAVD track is paired with another SAVD track and is assumed to be either a kaon or a pion. 
Thus each pair contributes twice in the combinatorial invariant mass distribution.
The combinatorial background is several orders of magnitude higher than the $D^0 + \overline{D^0}$ signal 
due to the low yield of charm particles. 
Five cuts were applied to reduce the large background. 
The cut parameters were chosen to maximize the signal-to-noise ratio (SNR) of the 
reconstructed $D^0 + \overline{D^0}$ peak and were determined from 
the \GeantFour-based simulations.
These cuts are:

\begin{enumerate}[(i)]
\item cut on the track transverse momentum, $p_T>0.34$ \GeVc;
\item cut on the track impact parameter, $d>37$ $\upmu$m;
\item cut on the longitudinal distance between the $D^0$ decay 
vertex candidate and the primary vertex, $V_\mcoor{z} > 1050$ $\upmu$m;
\item cut on the impact parameter D of the back extrapolated $D^0$ candidate momentum vector, D $< 18$ $\upmu$m;
\item cut on daughter tracks distance at the closest proximity, DCA $< 36$ $\upmu$m. 
\end{enumerate}
The $d$ and D parameters are defined as the shortest distance between the primary vertex and the track line
of a single track and $D^0$ candidate, respectively.
Note that the last four cuts are based on information delivered by the SAVD.

\begin{figure}
\resizebox{\linewidth}{!}{
\includegraphics{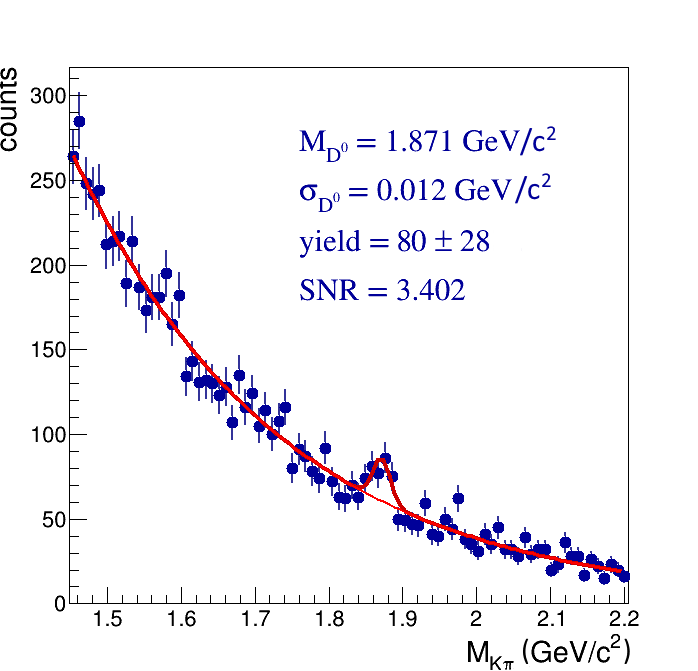}
}
\caption{Invariant mass distribution of unlike charge sign $\pi, K$ decay track candidates 
  for Xe+La collisions at 150\AGeVc taken in 2017. 
  The presented results refer to 1.86M 0--20~\% central events.}
\label{fig:Figure13}
\end{figure}

Fig.~\ref{fig:Figure13} shows the invariant mass distribution of unlike charge 
daughter candidates with the applied cuts for 1.86M 0--20\% central Xe+La events.
One observes a peak emerging at 1.86~\GeVcc, consistent with a $D^0 + \overline{D^0}$ production.
The invariant mass distribution was fitted using an exponential function 
to describe the background and a Gaussian to describe the $D^0 + \overline{D^0}$ signal contribution. Both lines representing signal plus background and background alone are drawn on the plot in red.
The indicated errors are statistical only.
From the fit, one finds the width of the peak to be 12~$\pm$~3.5~\MeVcc, consistent with the value 
obtained in simulations taking into account instrumental effects.
The total yield amounts to 80 $\pm$ 28 with a $\pm 3 \sigma$ integrated SNR of 3.4.  
The feasibility of $D^+$ and $D^-$ measurements has been demonstrated so far only by simulations.
However, these measurements are foreseen not to be more difficult than 
these of 
$D^0 $ and $\overline{D^0}$.

\subsection{$K^0_S$ and $\Lambda$ in the Xe+La data}
\label{sec:perf_xe+la}

\begin{figure}
\resizebox{\linewidth}{!}{
\includegraphics{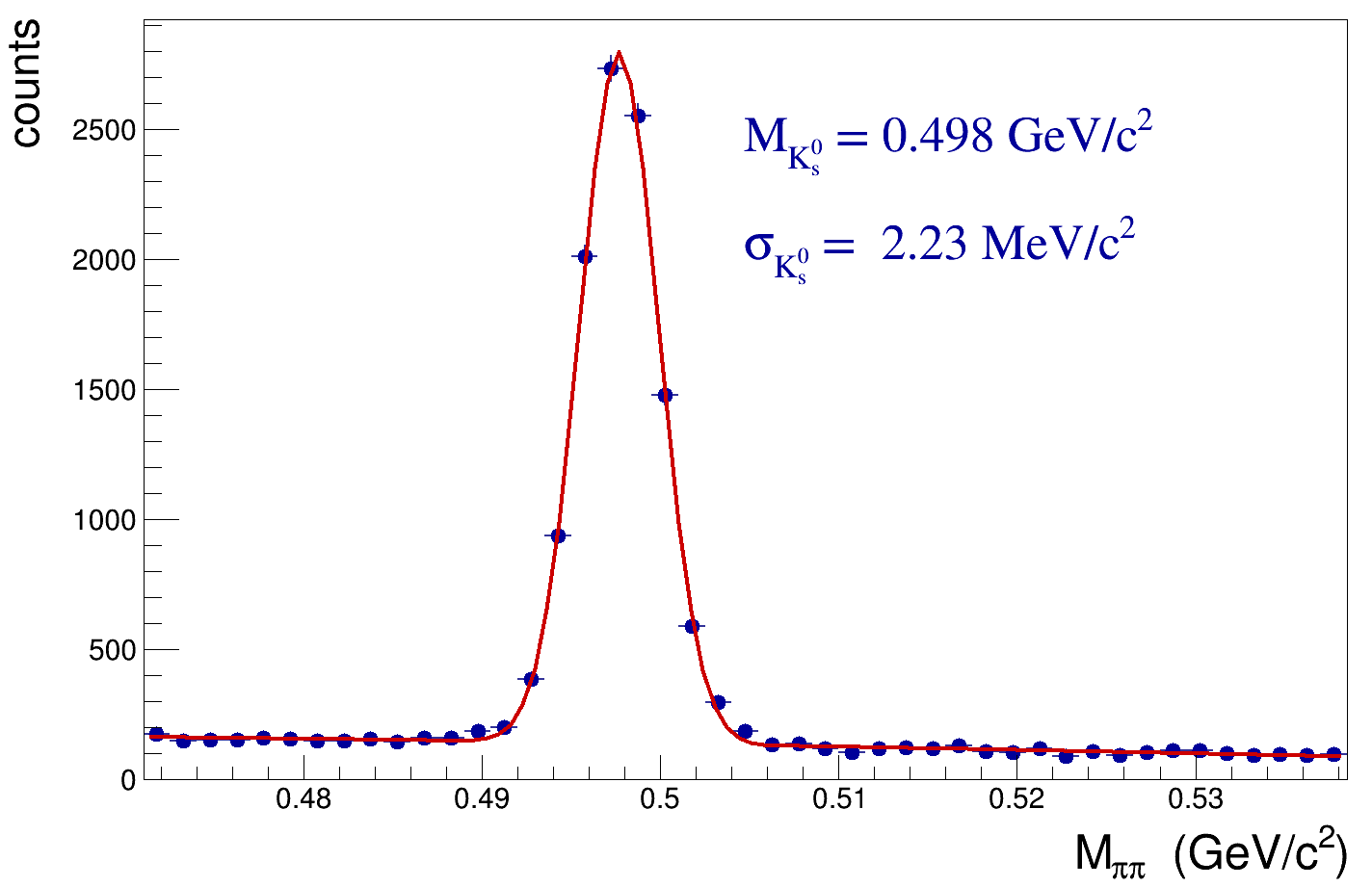}

}
\caption{Invariant mass distribution of unlike charge sign $\pi, \pi$ decay track candidates for Xe+La collisions at 150\AGeVc. 
The plot was done for 1.86M 0--20~\% central events.}
\label{fig:19}
\end{figure}

\begin{figure}
\resizebox{\linewidth}{!}{
\includegraphics{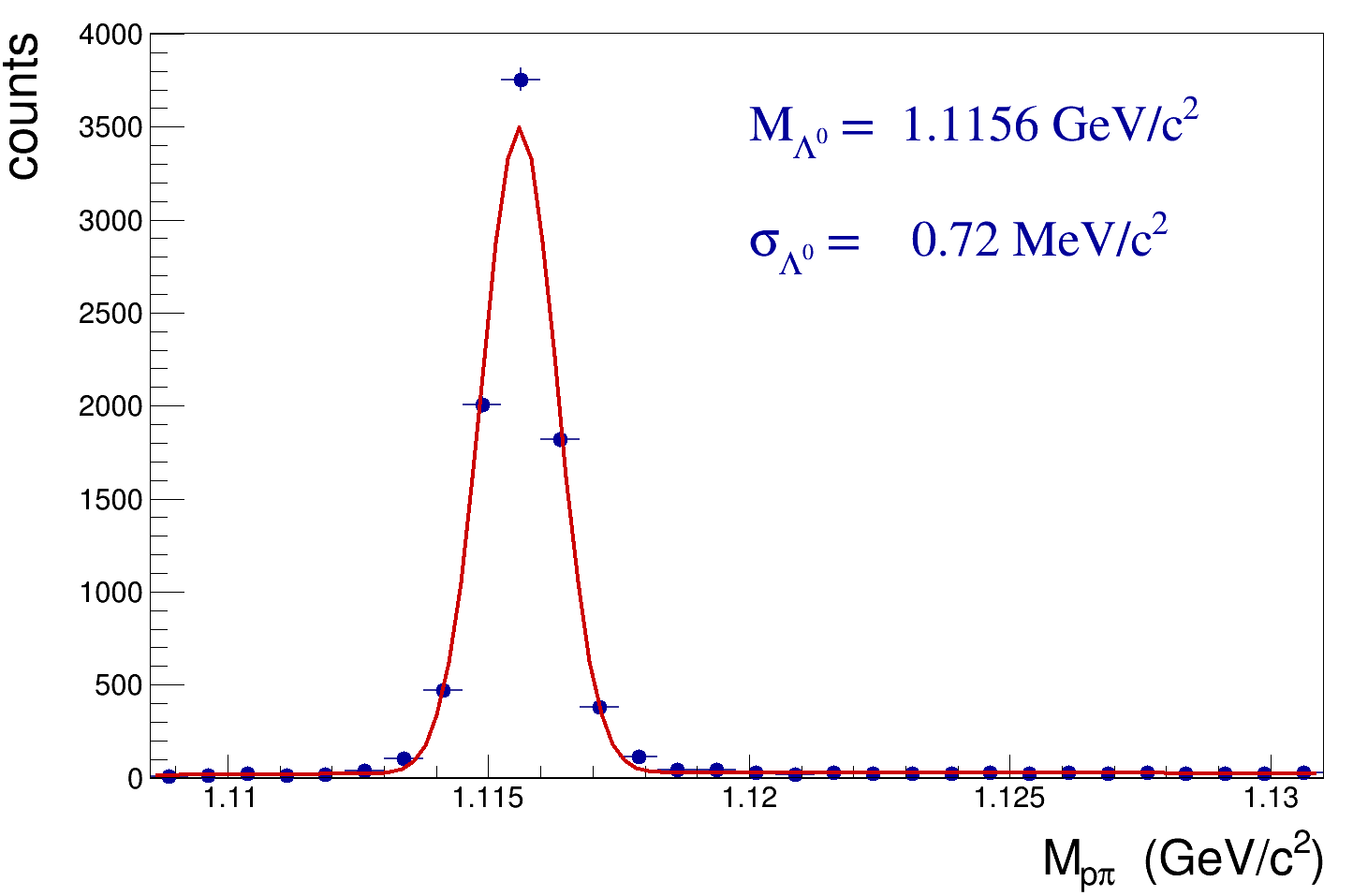}

}
\caption{Invariant mass distribution of unlike charge sign $\pi, p$ decay track candidates for Xe+La collisions at 150\AGeVc. The plot was done for 1.86M events of 0--20~\% central collisions.}
\label{fig:20}
\end{figure}

The same strategy of background suppression as that described in the previous section
can be applied for the reconstruction of $K^0_S$ and $\Lambda$ particles.
Fig.~\ref{fig:19} shows the invariant mass distribution in the regions of the $K^0_S$ mass
of the unlike sign pairs assigning $\pi$ mass to both tracks in the pair.
The results are drawn for $1.1 \times 10^6$ collisions of Xe+La at the beam momentum of 150\AGeVc. 
No event selection was applied. A clear $K^0_S$ peak is seen at 
0.498~\GeVcc.
For the same data
Fig.~\ref{fig:20} presents invariant mass distribution in the regions of the $\Lambda$ mass
for the unlike sign pairs assigning the proton mass to positively charged track and
the $\pi^{-}$ mass to negatively charged track in the pair.     
As in the case of $K^0_S$, a clear $\Lambda$ peak appears at the mass of 
1.1156~\GeVcc.
In both figures, the red line represents a fit with the Gaussian function to account for the signal plus 
the second-order polynomial to account for the remaining background.
The cut parameters were not optimized to maximize the signal significance. In this analysis, we used rather 
arbitrary cuts to demonstrate the ability of $K^0_S$ and $\Lambda$ reconstruction. 
As expected, the $\Lambda$ peak width is significantly smaller than the width of the $K^0_S$ peak. 
Our reconstruction over-predicts masses of $K^0_S$ and $\Lambda$ by 2~\MeVcc and 0.7~\MeVcc, respectively.
Although the shifts are small, they are much larger than the statistical uncertainty and are related
to the limited control of the absolute value of the magnetic field. 
The observed discrepancy can be used to calibrate the absolute strength of the magnetic field.  

\section{Summary and outlook}
\label{sec:6}

This paper presents the design and construction of a Small Acceptance Vertex Detector developed within \linebreak \NASixtyOne at the CERN SPS for pioneering measurements of open charm production. Moreover, the SAVD data calibration, event reconstruction and analysis procedure are also presented.

The SAVD was successfully operated at the top SPS energy during the test data taking on Pb+Pb collisions in 2016, Xe+La collisions in 2017, and Pb+Pb in 2018.
The recorded data allowed us to test the SAVD performance and develop the reconstruction procedures. The data analysis showed the track reconstruction efficiency and the reconstruction resolution of primary and secondary vertices, which are sufficient for measurements of open charm production within \NASixtyOne~\cite{Merzlaya:2021kue}.

Based on the experience gained with SAVD, the upgrade of the \NASixtyOne
the detector was performed during the CERN Long Shutdown~2, aiming for
accurate measurements of open charm. 
Most importantly, the data-taking rate was increased to 1~kHz~\cite{NA61Proposal}, and the new vertex detector has significantly improved performance.
It has 16 modules equipped with ALPIDE sensors developed within the ALICE ITS project~\cite{Aglieri:2013xma}.
This results in an enlarged acceptance of the device and increases the data-taking rate.
Recording of data on Pb+Pb collisions started in 2022 and
will continue over the Run~3 period.
The expected high statistics data should allow for 
detailed study of the open charm production in heavy ion collisions at the SPS energies.

\section{Acknowledgment}
\label{sec:7}
This work was supported by the Polish National Center for Science (grants
2018\slash29\slash N\slash ST2\slash02595, 2014\slash 15\slash B\slash ST2\slash 02537,
2015/18/M/ST2/00125, and 2018\slash 30\slash A\slash ST2\slash 00226),
the Norway Grants in the Polish-Norwegian Research Programme operated by
the National Science Centre Poland (grant 2019/34/H/ST2/00585), 
the German Research Foundation DFG (grant GA\,1480\slash8-1) and
St. Petersburg State University (research grant ID:75252518).\\
The results presented in this paper were obtained before February 2022.

\end{document}